# FEcMD: A multi-physics and multi-scale computational program for electron emission characteristics dynamically coupled with atomic structure in metal nano-emitters

Nan Li [a*], Xinyu Gao [b], Xianghui Feng [a], Kai Wu [a], Yonghong Cheng [a*] and Bing Xiao [a*]

[a] School of Electrical Engineering, State Key Laboratory of Electrical Insulation and Power Equipment, Xi'an Jiaotong University, Xi'an 710049, China.

[b] Xi'an Aeronautics Computing Technique Research Institute, AVIC, Xi'an Shaanxi 710068, China

[*] Corresponding author, E-mail: ln906061119@stu.xjtu.edu.cn, cyh@mail.xjtu.edu.cn, bingxiao84@xjtu.edu.cn

**Abstract**

Field emission coupled with molecular dynamics simulation (FEcMD) software package is a computational tool for studying the electron emission characteristics and the atomic structure evolution of micro- and nano-protrusions made of pure metals or multi-component alloys by means of multi-physics and multi-scale methodology. The implementations of molecular dynamics, the electrodynamics, and the heat conduction in FEcMD program are addressed. For molecular dynamics simulation, the Lennard-Jones potentials, embedded atomic method (EAM), and moment tensor potentials (MTP) are fully supported for both alloys and pure metals. In the electrodynamics, the FEcMD program incorporates the space charge fields (space charge potential and exchange-correlation effects) in the Wentzel-Kramers-Brillouin-Jeffreys (WKBJ) approximation to evaluate the field emission current density more reliably for nano-gaps between two metal electrodes. Additionally, the advanced two-temperature heat conduction model is implemented in FEcMD program, and which provides more reliable descriptions for the temperature evolutions of electron and phonon subsystems under the radiofrequency (RF) or pulse electric fields. Comprehensive benchmark tests are performed for each module in FEcMD software to validate the numerical results, and also to access the accuracy and efficiency of the implemented algorithms. Finally, some typical applications of FEcMD program are also demonstrated for understanding the evolution of temperature and electric field coupled with the dynamic changing of atomic structures for metal micro- and nano-protrusions.

**Program summary**

*Program* Title*:* FEcMD

*CPC Library* link *to program files:*

**1. Introduction**

In the modern era, electron field emission plays a crucial role in various applications, such as electron microscopies [1, 2], X-ray tubes [3], high power microwave sources [4], flat-panel displayers [5, 6], MEMS systems [7], electron lithography [8, 9], fusion reactors [10] and particle accelerators [11, 12], etc. Under the high electric field environment, field emission could trigger the thermal runaway event and eventually led to the development of a vacuum arc, resulting in the electric breakdown and other irreversible structural damages to the field emission devices or power equipment. Although, the vacuum electric breakdown has been linked to the structural damages at the atomistic scale and the associated thermal runaway process for a long time, a detailed coevolution of field emission characteristics and atomic structure changes under the high E-field and strong resistive heating is still not fully understood. In the past several decades, many experimental studies have been carried out to reveal the dynamic evolution of microstructure or atomic structure of field emitters in terms of micro- or nano-protrusion on the metal electrode surfaces under high E-field strength [13-15].Specifically, transmission electron microscopy (TEM) and scanning electron microscopy (SEM) could provide the ability to in-situ observe the deformation and structural phase transition of the micro- or nano-emitter before and after the vacuum electric breakdown [16-19]. However, neither TEM nor SEM can track the microstructure changes of field emitters during the electric breakdown or at the pre-breakdown moment because of the intense electromagnetic interferences. Otherwise, the vacuum breakdown usually occurs in an extremely short time scale, it is impossible to observe a transient phenomenon such as the atomic-scale breakdown process experimentally.

Computation simulations based on molecular dynamics simulations and finite element numerical calculations are considered as the alternative methodologies to investigate the strong coupling of microstructure evolution of field emitters with the E-field, heating, and electric stress at different stages of electric breakdown [20, 21]. In a typical finite element simulation, the fixed boundary conditions are employed to a specified geometry of micro- or nano-protrusions under the high electric field. The electron emission process is usually described by the classic Fowler-Nordheim equation or semi-classic quantum mechanical model [22] after knowing the local electric field distribution by solving either the Poisson equation or Laplace equation. In addition, the Joule and Nottingham heating mechanisms in the interior of electron emitters can be investigated through the heat diffusion process. Nevertheless, using the finite element techniques alone obviously lacks the ability to properly address the structural deformation, phase transition and thermal evaporation of micro- or nano-emitters under the influences of strong heating and high electric field stress. Among many recent developed numerical methods, Djurabekova [23-25] and coworkers proposed a multi-scale and multi-physics methodology known as the electrodynamics coupled molecular dynamics simulation (ED-MD). The ED-MD methodology has the unique feature that a continuous finite element description of spatial electric field distribution is dynamically coupled with the atomic structure evolution of field emitter, providing the ability to quantitatively understand the correlation between electric pre-breakdown and microstructure damages caused by the intense field emission process. More recently, the original ED-MD algorithm has been further developed to fully incorporate the three-dimensional (3-D) space charge effect on the local electrical field distribution. In the upgraded simulation, particle-in-cell (PIC) simulation of space-charge [26] was amended to ED-MD methodology, leading to a highly efficient and powerful computer simulation method which directly couples the finite element electrodynamics, particle-in-cell simulation, and discretized molecular dynamics simulation all together under two theoretical framework. Using ED-MD-PIC method, Veske and coworkers for the first time characterized the structural evolution of a conic Cu nanotips in multiple successive thermal runaway events under the high E-field [26]. Previously, we have also applied both ED-MD and ED-MD-PIC techniques to understand the structural damages and pre-breakdown electric properties of metal nanotips [27-30]. Nevertheless, the current implementations of ED-MD or ED-MD-PIC methodology showed some limitations, and which require further modification and extension of the algorithm to properly address the following aspects.

First, vacuum electric breakdown under the radiofrequency (RF) electromagnetic field and pulse voltage may also follow the cathodic model [31]. In the cathodic model, the electric breakdown mechanism is very similar to that of ordinary vacuum breakdown applying a static external E-field, i.e., the initiation of plasma is triggered by the intense electron emission and the thermal evaporation of micro-protrusions. However, the time scale of electron emission and the associated heat conduction process is usually proceeded at nanosecond or hundreds of ps range, and which is expected to be close to the typical RF or pulse duration. Under such circumstances, the thermal conduction mechanism in the interior of micro-protrusion or nano-protrusion maybe different to that of steady-state heating situation under a static E-field. More specifically, thermal energy dissipation should be described by solving two coupled heat balance equations for electron and phonon systems spontaneously. The phonon temperature is correlated to that of electron subsystem through the electron-phonon coupling constant which characterizes the thermal energy exchange rate between them. This heat transfer mechanism is known as the two-temperature model (TTM), and which has been widely adopted to simulate the damages in the materials caused by the heavy ion irradiation [32-34]. Recently, Uimanov and coworkers [35] implemented a two-temperature finite element simulation to investigate the electrode temperature evolution and the RF vacuum electric breakdown characteristics of micro-protrusions. It has been revealed that phonon and electron thermal conductions should be treated separately for the pulsed heating process at nanosecond range mainly because the thermal energy exchange between phonons and electrons is regulated by the electron-phonon coupling constant. To be more specifically, the electron temperature and phonon temperature must be obtained from the two coupled non-linear thermal conduction equations. For nano-emitters or nano-protrusions, the electric pre-breakdown could occur even at a shorter time scale than nanosecond [27-30, 36, 37]. Therefore, the realization of two-temperature model in ED-MD-PIC methodology certainly can provide more reliable physical insights regarding the thermal processes in the nano-emitters under RF or pulse voltage.

Second, for intense electron emission process in nano-gaps, the space charge density can be very high. As a result, the de Broglie wavelength and electrostatic potential energy of electrons are comparable to gap size and space charge energy, respectively. Under such conditions, electrons must be treated as the quantum particles, implying that the space charge potential and exchange-correlation potential of space charge (electrons) cannot be ignored when computing the space charge effects on the local E-field distribution and electron emission process. The quantum effects of space charge on the field emission current have been theoretically addressed by Ang and coworkers [38, 39] using a mean-field quantum mechanical model based on either free electron Thomas-Fermi approximation or the Kohn-Sham density functional theory (DFT). Specifically, the quantum Child-Langmuir law was derived for field emission current density in the cases of 1D and 2D nano-gaps [38]. The existing ED-MD-PIC methodology [20, 21, 27, 28, 40-42] does not consider the space charge fields (space charge potential and exchange-correlation effects) among emitted electrons in calculating the electron transmission coefficient using JWKB approximation in Kemble formula. Although, space charge fields are negligible for micro-size gaps applying to typical filed emitters, the including of space charge fields in the electron emission model is anticipated to significantly improve the accuracy and reliability of ED-MD-PIC simulations for electrodynamics and structural evolution when the spacing the between cathode and anode is in the submicron- or nano-range.

Finally, besides the pure elements, multi-component alloys are also widely used to fabricate the high-performance field emitters and power equipment. The vacuum electric breakdown mechanism of alloys initiated by the intense electron emission and thermal evaporation may be similar to that of single element case. Nevertheless, the current version of ED-MD-PIC program does not provide the ability to describe the structural evolution and thermal runaway process of alloys under the high E-field. Regarding the inter-atomic potentials, the embedded atomic method (EAM) employed in the ED-MD-PIC simulation is highly accurate for describing the structural properties of pure elements. However, fitting the EAM potentials for multi-component alloys remains a difficult task because the total number of different atomic pair types presented in the alloys increases rapidly with the number of elements. A large number of training data sets including the atomic structures, thermophysical properties, mechanical properties and thermodynamic properties must be prepared using the more accurate first-principles methods for all relevant binary atomic pairs. Nevertheless, machine learning potential (MLP) is considered as an alternative method to generate the inter-atomic potentials for alloys with high credibility. Specifically, the typical procedure to generate MLPs only requires atomic structures, energies, atomic forces, and stress tensors of snapshots from first-principles molecular dynamics (FPMD) simulation as the main input information for multi-variable regression algorithm [43]. Training the MLPs on-the-fly using the structural information, forces and energies from molecular dynamics simulation has been demonstrated in Appendix E. Furthermore, we have validated the elastic constants, melting point, lattice constants, and other properties of the copper, and found that they are in good agreement with experimental results. Interfacing the ED-MD-PIC algorithm to support both EAM potentials and MLPs for multi-component alloys is critical to improve the versatility and flexibility of the methodology, enabling the methodology to be used either for studying the atomic structure evolution of alloyed micro- and nano-emitters under the intense heating and high E-field or optimizing the field emission properties of nano-emitters made of multi-component novel alloys or the high-entropy alloys.

In this paper, we have made some significant upgrades in the algorithm of current ED-MD-PIC methodology to properly address the three aspects mentioned before. Firstly, the two-temperature model was implemented in the ED-MD-PIC simulation to provide a more reliable description for electron and phonon heat conduction processes under the RF electromagnetic field or pulse voltage. Notably, the two-temperature model enables the theoretical simulation to capture the atomic structure evolution of electron emitters within the high-frequency electronic heat conduction mechanism [35]. Secondly, the electron field emission model based on Wentzel-Kramers-Brillouin-Jeffreys (WKBJ) approximation and Kemble formula was modified to include the space charge fields including space charge and exchange-correlation potentials. Incorporating space charge fields would allow ED-MD-PIC algorithm to be more reliable for predicting the emission current density and local E-field distribution on nano-emitter whenever the space charge density is high, and electrons should be treated as quantum particles. It is also worth mentioning that the LIBXC library was interfaced with the algorithm, enabling the use of exchange-correlation functionals going beyond that of local density approximation (LDA) [44, 45]. Thirdly, the numerical algorithm relevant to molecular dynamics simulation in current ED-MD-PIC methodology was completely rewritten, fully supporting three different types of interatomic potentials, i.e., the Lennard-Jones pair potential (LJ), EAM potentials and machine learning potentials (MLPs) in terms of moment tensor potentials (MTP) for alloys [43]. All new upgrades together with other key external open-source libraries (FEMOCS [21] and GETELE [40]) employed in the current ED-MD-PIC program form the new computational tool that is called "Field Emission coupled Molecular Dynamics" or "FEcMD" in this work.

The present paper is organized as follows: The overall workflow of FEcMD program is illustrated in section 2. In section 3, the theoretical backgrounds and the implementation details are presented. In section 4, we demonstrate some interesting applications of FEcMD program to model the field emission and atomic structure of micro- or nano-protrusions or nano-emitters under the high static and RF E-fields.

**2. Workflow of FEcMD simulation**

The FEcMD software consists of four principal computational modules, including molecular dynamics (MD), atomic forces (AF), electrodynamics (ED), and heat conduction (HC), as illustrated in Figure 1. The additional finite-element mesh (FEM) generation module mainly serves for HC and ED modules to solve the heat balance equation for heat conduction and Poisson equation for electrodynamics. The FEM module also plays the key role in FEcMD software to dynamically couple the variation of geometry of nanotip obtained from MD module with HC and ED modules that determine the temperature and E-field distributions, supporting the multi-scale and multi-physics simulations for nano-emitters using ED-MD-PIC methodology. To perform the standard ED-MD-PIC simulation, all computational modules presented in Figure 1 must be employed. Otherwise, another useful feature of FEcMD software is that users can also use MD+AF, HC+FEM, ED+FEM as the stand-alone computational tools to perform the molecular dynamics simulation, heat transportation and electrodynamics, respectively. Besides the ED-MD-PIC simulation, the use of other additional stand-alone features requires no further modification of the FEcMD software, and only a minor change in the input file is needed. Nevertheless, the rigid shape is assumed for the nano-emitters in calculating the temperature and E-field evolutions within the stand-alone HC+ED method using finite element method. Meanwhile, the MD+AF requires no finite element mesh generation at all in the molecular dynamics simulation. From Figure 1, it is easy to see that the HC module perturbates the atomic velocities in MD simulation, and ED module contributes additional interatomic and external forces to the AF module that are eventually combined with the intrinsic interatomic potentials to give rise the net forces acting on atoms of nanotip during the MD calculation. The atomic structural evolution changes the shape and size of nano-tip, directly resulting in a dynamic coupling of atomic structure with the temperature and E-field distribution through a self-adaptive updating of the finite element mesh in the ED-MD-PIC simulation.

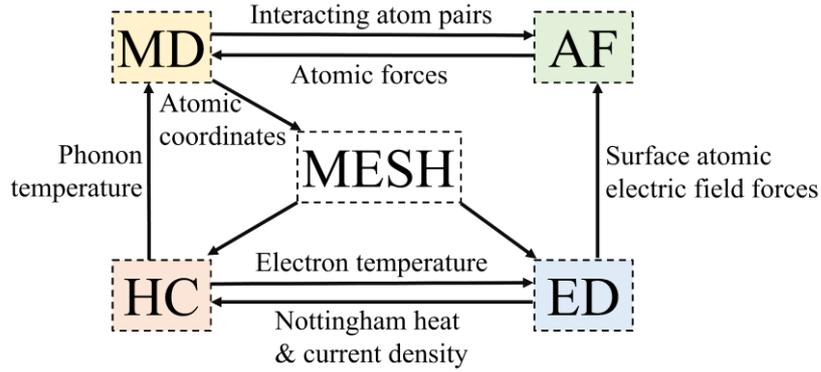

**Figure 1.** Principal computational modules in the framework of FEcMD software and their feedback loops.

In Figure 2, the detailed workflow of FEcMD software is illustrated. The MD module together with AF module could perform the standard molecular dynamics simulations at the finite temperature under various thermodynamic ensembles (NVT, NVT, NPT and non-equilibrium MD) without coupling to ED and HC modules. In the current implementation, the MD module provides the interface to conduct the structural evolution using Lonnard-Jones (L-J) pair potential, embedded atomic method (EAM) potential, and moment tensor potential (MTP). The AF module is responsible for evaluating both the intrinsic inter-atomic forces and external disturbing forces (Lorentz force) for atoms. Notably, when performing the standard MD simulation independently, the external atomic forces such as Lorentz force and Coulomb force for charged ions on the nanotip surface are not calculated. In the AF module, the realization of MLP is supported by the open-source MLIP library [43]. The electrostatic potential and local electric field distributions are obtained from ED module where the Poisson equation is solved numerically in three-dimensional (3D) space on a finite element grid using FEMOCS library [21]. Electron emission current density distribution and total emission current density are calculated using GETELE library in ED module. The space charge density distribution is determined from PIC simulation in ED module. The space charge potential and exchange-correlation potential of space charges are obtained from our new implementations and the open source LIBXC library. Then the WKBJ approximation and Kemble formula are employed to calculate the electron transmission coefficient and electron emission current density. It is worth mentioning that the algorithm implemented in GETELE library is capable of determining the electron emission current density reliably from a combined thermal-field emission mechanism [40]. After knowing the electron emission current density distribution on the emitter surface, the Nottingham heat is obtained from GETELE library. Besides the static E-field, the ED module in FEcMD software also provides the ability to support the ED-MD simulations under the RF and pulse voltages. The electric current density and the associated resistive Joule heating are treated by HC module. In the HC module, the temperature evolution in the interior of electron emitter is calculated by numerically solving the heat balance equation on a finite-element mesh. The finite-element mesh generation is supported by the open-source Deal II library in vacuum region, surface, and the interior of electron emitter [46]. The extraction of surface atoms is realized by performing the atomic coordination number analysis within the density-based spatial clustering of applications with noise algorithm (DBSCAN). The temperature distribution is transformed into the velocity disturbances to atoms in MD simulation. Otherwise, both MD and HC modules provide the post data processing functionalities for structural properties and filed emission characteristics including the atomic coordination number (ACN), atomic radial distribution function (RDF), atomic short-range order parameter (SRO), field emission current density distribution, E-field distribution, electron/phonon temperature distribution. The processed data can be further analyzed and visualized by the users using other graphic programs such as OVITO [47], ParaView [48], VMD [49], and VESTA [50].

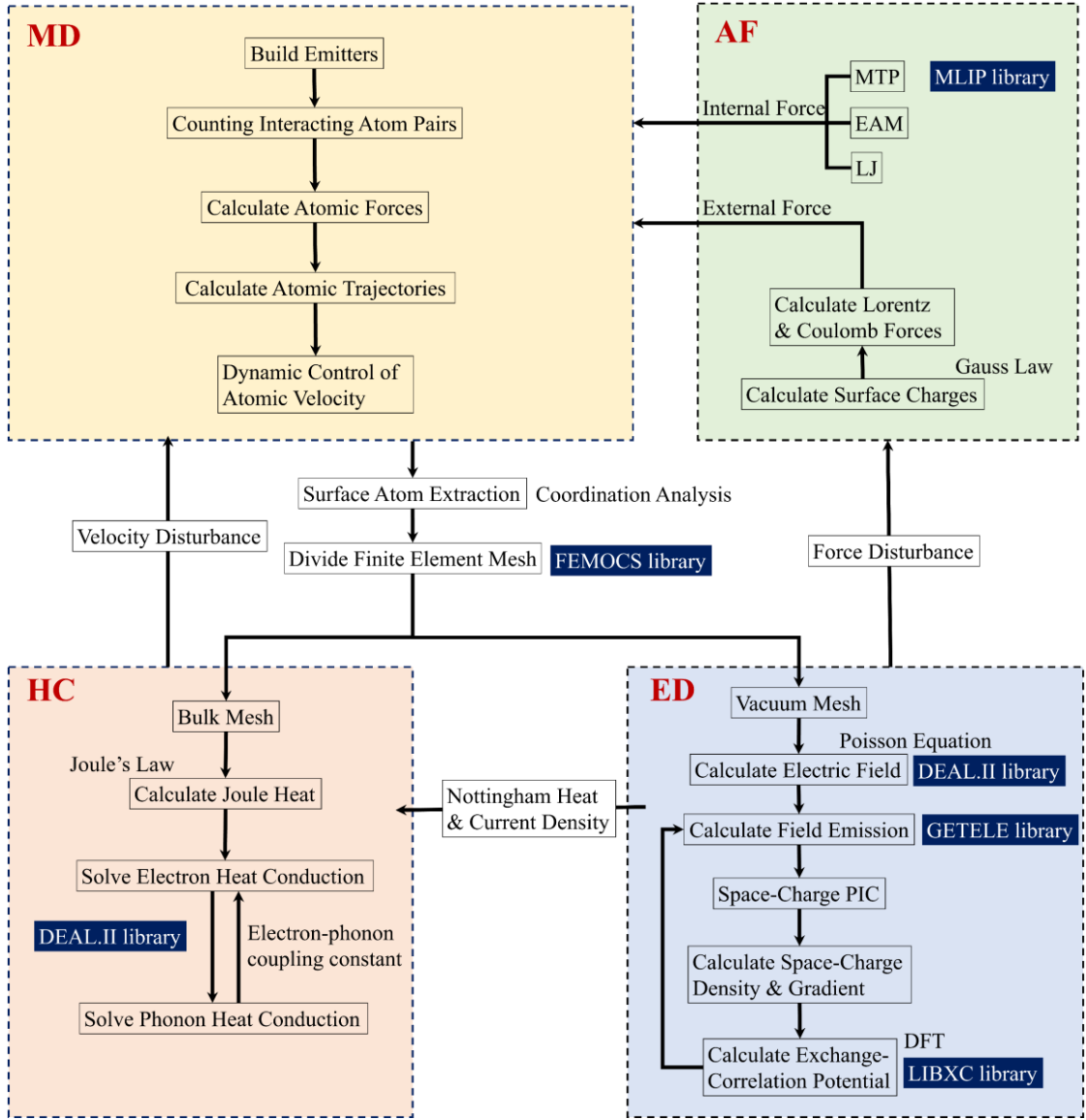

**Figure 2.** Workflow of FEcMD software.

## 3. Methods and Algorithms

### 3.1 Molecular dynamics (MD) simulation

FEcMD software package employs C++ to implement all algorithms related to classic molecular dynamics simulations. The fundamental principles and numerical algorithms of classic MD simulation can be found in Ref [51], including the realization of thermodynamic ensembles, atomic position and velocity integrations, and statistics for structural and physical properties, as shown in Figure 3. The solid lines highlight the MD and AF modules, which are required for running molecular dynamics (MD) simulations independently. The MD module calculates atomic accelerations, velocities, and then atomic coordinates are renewed according to Newton's second law. Meanwhile, the AF module provides the net force acting on each atom in nano-emitters. Some key techniques employed in the MD module are worth mentioning here.

- To achieve the high computational efficiency for a large system containing more than $10^5$ atoms, the inter-atomic forces are obtained using the neighbor-list algorithm [51]. This is achieved by updating the neighbor lists within a user defined cutoff radius ($\Delta r$) and time interval ($\sum(\max|v_i|) > \Delta r/2\Delta t$).

- Both the leapfrog and prediction-correction methods are implemented for particle position and velocity integrations. Leapfrog integration method demands minimal storage space and less numerical calculation, and which is an optimal option for large-scale MD simulation. Otherwise, prediction-correction algorithm shows higher numerical accuracy than that of leapfrog method, and which suits for the rigid body and constrained MD dynamics.

- Both the equilibrium and non-equilibrium molecular dynamics simulations are implemented in the MD module. For equilibrium MD simulation, NVE, NVT and NPT ensembles are realized in the current algorithm. In the case of non-equilibrium MD simulation, the MD module performs structural relaxation within a directional temperature gradient, allowing users to calculate the lattice thermal conductivity of the materials using Fourier law. For equilibrium MD simulation, the three-dimensional periodic boundary conditions (PBCs) and Kubo-Greenwood formula are realized to calculate the lattice thermal conductivity.

- Supporting the state-of-the-art interatomic potentials especially the machine learning potentials to conduct the molecular dynamics simulations for multi-component alloys and compounds.

Three different inter-atomic potentials are fully supported by FEcMD software package in the AF module, including Lennard-Jones potential, EAM potential and MTP potential. L-J potentials have been widely used for performing classic MD simulations ever since the earlier days of computer simulation. The L-J potential is considered as the simplest pair potential, and which is given by Eq. (1). The use of L-J potential only requires two parameters, the $\sigma$ and $\varepsilon$, for an atomic pair. The L-J potential is regarded as a good approximation for describing the isotropic forces among particles, i.e., long-range van de Waals interactions. However, better potential is needed when inter-atomic interactions are strong and highly anisotropic.

$$E_{ij} = \begin{cases} 4\varepsilon\left[\left(\dfrac{\sigma}{r_{ij}}\right)^{12} - \left(\dfrac{\sigma}{r_{ij}}\right)^{6}\right] & r_{ij} < r_c \\ 0 & r_{ij} \geq r_c \end{cases} \qquad (1)$$

The EAM potential can accurately describe the many-body interactions of metals and alloys, and which is given by Eq. (2) [52]. As can be seen from Eq. (2), the EAM potential consists of two terms, the pair potential ($\phi_{\alpha\beta}$) and the embedding energy ($F_\alpha$). The EAM potential in FEcMD software package is realized using the open-source EAM potential library in LAMMPS code, mainly contributed by Foiles and coworkers [53].

$$E_i = F_\alpha\left(\sum_{j \neq i} \rho_\beta(r_{ij})\right) + \frac{1}{2}\sum_{j \neq i} \phi_{\alpha\beta}(r_{ij}) \qquad (2)$$

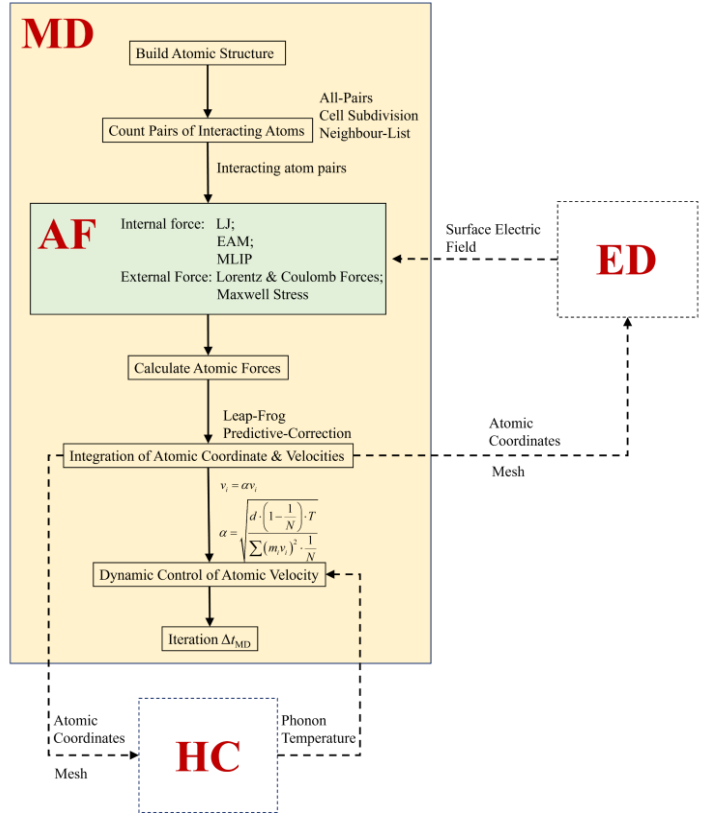

**Figure 3**. Algorithms and workflow of molecular dynamics simulations. The solid squares highlight the principal MD and AF modules, and other supporting modules in FEcMD software are indicated as the dashed squares.

Generating EAM potentials for novel alloys requires tremendous works because many structural, mechanical and physical properties of the material are needed for numerical fitting and subsequent validation procedures. Machine learning potentials can effectively mitigate the tedious numerical fitting procedure of EAM potentials. Among different machine learning potential forms, the moment tensor potential (MTP) is adopted in our FEcMD software package, and which is given by Eqs. (3) and (4) [43]. MTPs use the basis functions ($B(\boldsymbol{n}_i)$) to calculate the interatomic potential energy ($E_i$). The exact form of a basis function is determined from the moment tensor descriptors or also known as moments ($M(\boldsymbol{n}_i)$). Each moment has the radial and angular parts as two major components in its construction (See Eq. (4)). Eventually, the functional form of MTP is controlled by two parameters, the level of moments and the size of radial basis (polynomial functions). The main advantage of using MTP is that it could accurately describe an arbitrary local atomic coordination environment using the linear regression algorithm for a set of basis functions. Otherwise, MTPs also support the MD simulation for multi-component alloys.

$$E_i = \sum_{level} \xi B(\boldsymbol{n}_i) \tag{3}$$

$$M(\boldsymbol{n}_i) = \sum_j f(r_{ij}, \alpha, \beta) \boldsymbol{r}_{ij} \otimes ... \otimes \boldsymbol{r}_{ij} \tag{4}$$

Fitting MTPs require the energies, forces, and stresses from multiple atomic configurations as the training set. The most convenient way to generate those training data sets is to perform molecular dynamics simulations based on first-principles method or force field approach. The external computational tools such as VASP and LAMMPS could be used for the purpose. The realization of MTPs in FEcMD program is achieved by using the open-source machine-learning interatomic potential (MLIP) package [43]. A standard input file for fitting MTPs using MLIP library in FEcMD software package is given in Appendix B. In section. 4.4, we also demonstrate the creation and the use of MTPs for Cu and W-Mo alloys.

In addition to calculate the intrinsic inter-atomic forces, the AF module also utilizes Gauss's law to calculate the charge carried by the surface atoms [20]. Then, the charged surface atoms experience both inter-atomic Coulomb forces [25] and Lorentz forces under the E-field in ED-MD simulation. The total charges of partially ionized surface atoms and the Lorentz forces are obtained from Eqs. (5) and (6).

$$q_i = \sum_{f_{MV}} \overrightarrow{F_f} \cdot \hat{n}_f A_f \tag{5}$$

$$\overrightarrow{f_L} = \frac{1}{2} q_i \overrightarrow{F_i} \tag{6}$$

Here, $F$ refers to the electric field strength, and the term " $1/2$ " implies that only one side of the surface atoms, specifically the side in contact with the vacuum, is exposed to the electric field; The term "$q_i$" means the net charge on each ionized atom according to Gauss's law in Equation (5); "$f_{MV}$" signifies that for each finite element cell containing a surface atom *i*, and the integration is performed over the product of the local electric field at the cell face center (or the Voronoi cell facets) that is exposed to vacuum, the unit vector perpendicular to the cell face, and the area of the face.

Another interesting feature of the AF module is that it also includes a simple algorithm to compute the Maxwell stress for surface atoms in the case of a planar electron emitter. Previously, such algorithm has been applied to study the initiation and growth of nano-protrusions on planar polycrystalline Cu electrode under high E-field [54]. For the planar metal electrode, the Maxwell stress and the corresponding electric field forces acting on surface atoms are given by Eqs. (7) and (8).

$$\sigma = \frac{\varepsilon_0}{2} F^2 \tag{7}$$

$$\vec{f}_{ex} = \frac{\sigma S}{N} \vec{n} \tag{8}$$

Where, $\sigma$ denotes the Maxwell stress, and $F$ refers to the electric field strength. $S$ and $N$ represent the surface area and total number of surface atoms for planar electrode, and $\vec{n}$ is the surface normal. In the MD simulation, $f_{ex}$ is added to the interatomic forces obtained from interatomic potentials, resulting in the total atomic force. The determination of surface atoms is reliably done by conducting the atomic coordination number analysis (CNA).

In a routing ED-MD multi-physics simulation, the MD module receives a temperature disturbance from the HC module due to the Joule and Nottingham heating processes. The temperature disturbance is obtained by solving the heat balance equation in the interior of electron emitter on a finite element mesh. In MD simulation, velocities of constituting atoms in electron emitter must be regulated according to temperature profile determined from heat balance equation. The regulation is realized by discretizing the electron emitter into multiple stacking layers with a thickness $\Delta z$ for each layer. After solving the heat balance equation for electron emitter, the temperature evolution is found for each sliced layer. Then, the velocities of particles belonging to the same slice are scaled to match the macroscopic temperature profile. Rescaling the velocity of each atom is realized by Eq. (9) [51].

$$v_i = \gamma v_i \tag{9a}$$

$$\gamma = \sqrt{\dfrac{d \cdot \left(1 - \dfrac{1}{N}\right) \cdot T}{\sum (m_i v_i)^2 \cdot \dfrac{1}{N}}} \tag{9b}$$

Where, $\gamma$ refers to the rescaling constant, and $N$ denotes the total number of particles in a particular slice. For 3-dimenisonal MD simulation, $d$=3. Other symbols have their usual physical meanings.

**3.2 Electrodynamics (ED) module**

The ED module is written in C++, and which is directly interfaced with FEMOCS [21] and DEAL.II [46] libraries for solving Poisson or Laplace equation using finite-element method. Figure 4 illustrates the principal workflow of ED module. The ED module in FEcMD software package solves the Poisson or Laplace equation (See Eq. (10)) on the finite-element mesh to obtain the electrostatic potential and the E-field strength locally (See Eq. (11)). The ED module can be used as a stand-alone computational tool to obtain the spatial distribution of E-field and electron field emission properties of a nano-emitter with a rigid shape and fixed boundary conditions on finite element mesh. In order to generate the finite element mesh for the electron emitter, ED module implements an automatic workflow with the built-in scripts to define the size and shapes of a nano-tip. The users could simply use those scripts to create and edit the atomic structures of metal nano-tip. Then, the atomic structure of the nano-tip is adopted to define the interior, the surface, and the vacuum region in the finite element simulation based on the coordination number analysis. Separation of finite element meshes in three different regions is realized using the DBSCAN algorithm (Density-Based Spatial Clustering of Applications with Noise) [55]. Eventually, the open-source Deal II library solves the Poisson or Laplace equation on the finite element mesh. The emission current density is obtained for each finite element cell on the surface of nano-tip after knowing the local E-field value that is normal to the surface (Voronoi cell facets). The electron field emission current density on the surface of the nano-emitter is obtained using the GETELE library within the WKBJ model. Notably, the GETELE lib has already been integrated into the open-source FEMOCS library [40]. The space charge effects on the local electric field distribution are considered due to the use of particle-in-cell (PIC) simulation when solving Poisson equation. Employing the PIC method is critical in ED-MD-PIC methodology, because it directly provides the space charge density and its distribution in the vacuum region between anode and cathode. From the obtained space charge density distribution, the charge density gradient and the Laplacian are further calculated. The above information is needed to calculate the space charge potentials such as space charge potential and exchange-correlation potential in WKBJ model to obtain the electron tunnelling barrier profile on the finite element mesh at the surface of nano-tip. For conventional electron field emission configuration, the space charge has a relatively low density and large kinetic energy, implying that the emitted electrons can be treated as the classic particles. Under such circumstances, the quantum many-body effects of space charges are completely negligible. On the other hand, the space charge potential and exchange-correlation potential may play significant or even the dominant role in determining the overall electron tunnelling barrier profile for nano-gaps where field emission can reach either the classic space charge limited regime (the classic Child-Langmuir law) or the quantum electron field emission regime (the Child-Langmuir law in quantum regime)[37]. In the current workflow of ED module in FEcMD software, it is optional to include space charge potential and exchange-correlation potential when solving the WKBJ model to compute the electron tunneling barrier profile and field emission property. This feature provides additional flexibility for users to incorporate the appropriate field emission mechanisms with an optimal numerical efficiency in a computational recipe.

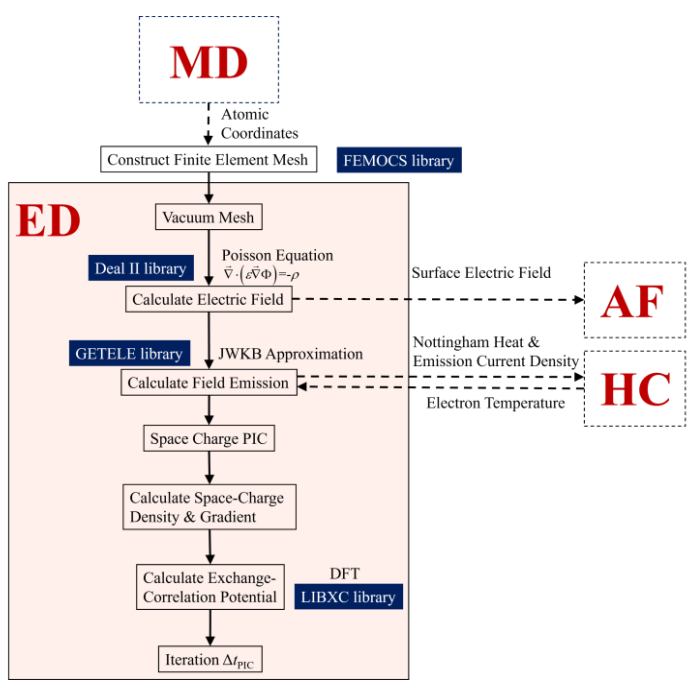

**Figure 4** Algorithms and workflow of electrodynamics module for calculating electric field evolution and electron field emission characteristics. The principal ED module is highlighted as the solid square. Other supporting modules are denoted by dashed squares including MD, AF and HC modules. The iteration time step of particle-in-cell (PIC) with electric field is given by $\triangle t_{PIC}$.

The boundary conditions employed for solving Poisson equation is illustrated in Figure 5. Obviously, the mixed Neumann and Dirichlet boundary conditions are used. As can be seen from Figure 5, the nano-tip is placed in a large rectangular box which is periodic in the lateral directions. In the vertical direction, the upper surface of the box represents the boundary of anode, and which is usually assumed to be sufficiently far away from the cathode below for the conventional electron field emission devices. Obviously, the anode is treated as a planar electrode with a fixed E-field value that matches the macroscopic the applied E-field (Neumann boundary condition). The surface of a metal nano-tip is equipotential, and the value is set to 0 V (Dirichlet boundary condition). For surfaces of the box in the lateral directions, the electric field in either x or y direction is also set to 0 V (Neumann boundary conditions). Meanwhile, the electrostatic potential in the vertical direction (or z-direction) is obtained from finite element method self-consistently. Both the height and lateral dimensions of the simulation box are significantly larger than that of nano-emitter in the typical field emission structure. In the case of nanogap, the vertical height of the box is determined by total height of nanotip and the applied macroscopic E-field value.

$$\nabla^2 \varphi = -\frac{e\rho}{\varepsilon_0} \tag{10}$$

$$E(r,z) = -\nabla \varphi(r,z) \tag{11}$$

Besides the static E-field, we also implemented algorithms to support the pulse and RF electric fields, as given by Eqs. (12) and (13). The pulse E-field is defined by the pulse duration ($t_{max}$) and pulse amplitude ($F_0$). On the other hand, the RF electric field is approximated by sine function with the user defined angular frequency ($\omega$) and amplitude ($F_0$).

$$F = \begin{cases} F_0, & 0 < t < t_{max} \\ 0, & \text{others} \end{cases} \tag{12}$$

$$\begin{aligned} F &= F_0 \sin(\omega t) \\ \omega &= 2\pi f \end{aligned} \tag{13}$$

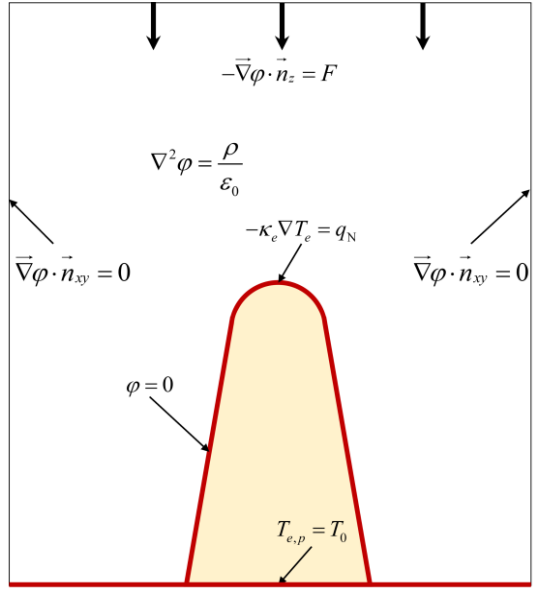

**Figure 5** Typical geometry for micro- or nano-protrusion, and the boundary conditions employed for ED-MD simulations in FEcMD software.

In FEcMD software, ED module interfaces directly with the GETELE library to calculate the field emission current density *J* and Nottingham heat $P_N$, which are expressed by Eqs. (14) and (15) [40, 56]. The electron transmission coefficient *D*(*E*) is obtained by WKBJ approximation and Kemble formula, as shown in Eqs. (16) and (17).

$$J = Z_s k_B T_e \int_{-\infty}^{\infty} D(E) \log\left(1+\exp(-E/k_B T_e)\right) dE \tag{14}$$

$$P_N = Z_s \int_{-\infty}^{\infty} \frac{E}{1+\exp(E/k_B T)} \int_{-\infty}^{E} D(\xi) d\xi dE \tag{15}$$

$$D(E) = \frac{1}{1+\exp(G(E))} \tag{16}$$

$$G(E) = g \int_{x_1}^{x_2} \sqrt{U(x)-E}\, dx \tag{17}$$

Where, the *D*(*E*) refers to electron transmission coefficient, and *G*(*E*) denotes the Gamow exponent. *U*(*x*) represents the electron tunneling barrier. $Z_s \approx 1.618 \times 10^{-4}$ A(eV nm)$^{-2}$, is the Sommerfeld current constant and $T_e$ is the electron temperature. In FEcMD program, the electron tunneling barrier *U*(*x*) is given as Eq. (18). The first three terms in Eq. (18) represent the work function, the applied external electric potential, and image charge potential. The fourth term accounts the quantum many-body interactions among the emitted electrons, and which is calculated from the exchange-correlation functional in density functional theory (DFT). The last term in Eq. (18) gives the space charge potential. The space charge potential and exchange-correlation potential could significantly affect the electron tunneling barrier $U(x)$ when the emission current density is high in the nano-gap [37, 39].

$$U(x) = \phi_{WF} - eV(x) - \varphi_{ic}(x) + \varphi_{xc}(x) + \varphi(x) \tag{18}$$

Regarding the image charge potential in Eq. (18), the classic expression is written as Eq. (19). The classic form of image charge potential has the singularity at $x = 0$, and calculated electron emission energy barrier profile is unrealistic at the very small distance near the interface of emitter and vacuum. Otherwise, Eq. (19) may only apply to the planar electron emitter.

$$\varphi_{ic}^{classic}(x) = -\frac{e^2}{16\pi\varepsilon_0 x} \tag{19}$$

For the non-planar emitter, Ref [57] suggested that Eq. (20) may be used to calculate the image charge potential by considering the local radius of curvature (RoC). However, Eq. (20) does not eliminate the singularity of image charge potential at the interface between electrode and vacuum ($x = 0$).

$$\varphi_{ic}^{classic}(x) = -\frac{Q}{16\pi\varepsilon_0 x(1+x/2R)} \tag{20}$$

To eliminate the singularity in image charge potential, a more advanced method must be employed. One such method is the use of free electron Thomas-Fermi approximation (TFA) with the random phase approximation (RPA) to calculate the image charge potential [58]. In this method, the image charge potential is obtained from Eq. (21). Here, the Green function of a longitudinal self-consistent field $D_{vac}(p, x, x')$ describes the screened Coulomb interaction between the charges at points $x$ and $x'$, and $p$ is the wavevector along the $x$ direction (field emission direction).

$$\phi_{ic}^{TFA}(x) = -\frac{e^2}{4\pi\varepsilon_0} \int_0^\infty p\,dp \left[ D_{vac}(p,x,x') + \frac{1}{2p} \right] \tag{21}$$

For the non-planar electron emitter, the local electric field employed in Fowler-Nordheim equation to calculate the emission current density must the corrected by considering the local geometry of nanotip. Previously, Kyritsakis and coworkers employed the local radius of curvature (RoC) as the main variable to correct the electrostatic potential for sharp electron emitters down to RoC of 4~5 nm [40]. In FEcMD software, solving Poisson or Laplace equation directly gives the local electrostatic potential values on the finite element mesh. Then, the cubic spline interpolation method is used to smoothly evaluate the electrostatic potential along the electron field emission direction on nanotip.

To obtain the space charge density for calculating the space charge potential, and exchange-correlation potential, we employed the PIC technique to simulate the emission, collision, and motion of electrons. For more technique details and algorithms, we refer to Ref [26].

In FEcMD software, we implemented Eqs. (14)~(21) for calculating the electron tunneling barrier and electron emission current density within the quantum effects and space charge potential of space charges in the modified open source GETELE library. In addition, the GETELE library [40] is also interfaced to the LIBXC library [44, 45]. LIBXC library provides hundreds of different semilocal exchange-correlation functionals and their numerical algorithms, and which greatly improves the computational efficiency of exchange-correlation potential for space charge in ED module. In previous studies on the quantum model of space charge limited current density in nano-gap, the local density approximation (LDA) has been widely used in Eq. (18). Other options for computing the quantum effects of space charges are also known, including the generalized gradient approximation (GGA) and meta-generalized gradient approximation (meta-GGA). Using LDA, GGA and meta-GGA functionals in Eq. (18) is convenient mainly because those semilocal exchange-correlation density functionals only rely on the space charge density ($\rho(r)$), the density gradient ($\nabla\rho(r)$), and the Laplacian ($\nabla^2\rho(r)$) in the numerical calculations. Those quantities are straightforward to compute on the finite element mesh after knowing the space charge distribution. More technique details and numerical results regarding the use of different exchange-correlation functionals to treat the quantum many-body effects of space charges can be found in our recent work and the reference therein [59].

**3.3 Heat conduction (HC) module**

The workflow of HC module is written in C++, and which interfaces with Deal II library for solving two-temperature heat conduction equations using finite-element method, as shown in Figure 6. We have implemented the coupled two-temperature model in the Deal II library. The ED and HC modules in FEcMD software share the same method and algorithms for constructing finite-element mesh using DUBSCAN algorithm. However, HC module only utilizes the finite element mesh grids in the interior of nano-emitter to numerically solve the heat balance equations for electron and phonon subsystems. It is in contrast to that of ED module where the electric potential and E-field are obtained in the vacuum region and mesh grids that are adjacent to surface atoms of metal nano-emitter. In FEcMD software, HC module can be used independently by the users to simulate the heat conduction and temperature evolution in the micro-protrusions and nano-emitters within the two-temperature model and a rigid geometry. Two-temperature model, which calculates the phonon and electron temperatures separately, is considered as a more accurate methodology to understand the electron field emission property and the associated heating process in micro-protrusions and nano-emitters under the time-dependent E-field such as pulsed voltage and radio frequency electromagnetic field. Otherwise, coupling molecular dynamics simulation with HC module keeps updating the finite element mesh for heat transfer calculation due the variations of the shape and geometry of electron emitter caused by mechanical deformation or melting due to the heating process. The velocities of atoms in the atomistic simulation are also regulated by the temperature distribution on the finite element mesh in electron emitter obtained from two-temperature heat conduction model. Heating of electron emitters (micro-protrusions or nano-tips) is attributed to the resistive heat (Joule heat) and field emission related heat deposition (Nottingham heating or cooling) processes. Those heating sources are invoked by the use of ED module together with HC module in FEcMD software.

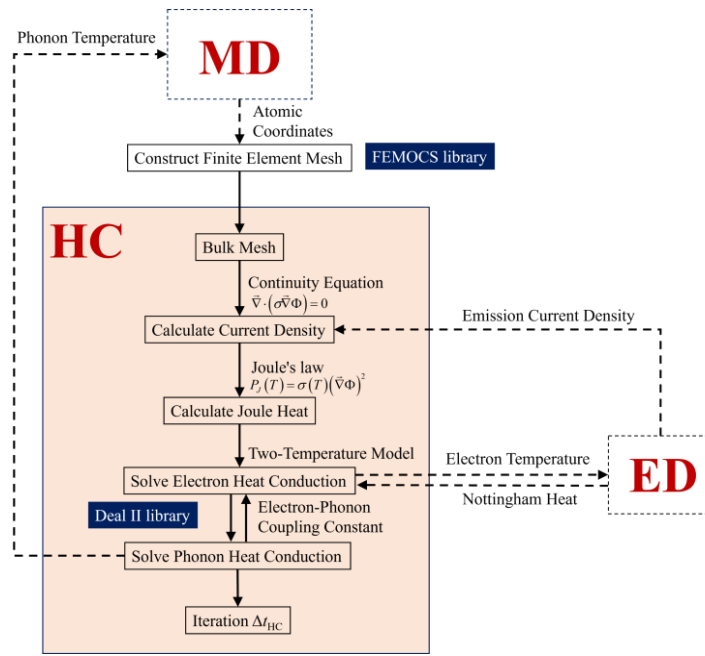

**Figure 6**. The algorithms and workflow of HC module for solving electron and phonon heat balance equations using the coupled two-temperature model on finite element mesh. Solid square represents the stand-alone HC module and dashed squares denote other computational modules that are also compatible with HC module. The iteration time step of heat diffusion refers to $\triangle t_{HC}$.

In the case of time-dependent E-field, the electron field emission, and the associated heating processes (Joule and Nottingham heating) in micro- or nano-protrusion could last few ns or hundreds of ps, depending on the frequency of RF electromagnetic field or the duration of pulse voltage (See Eqs. (12) and (13)). The heat conduction mechanism during a transient resistive heating process could be different to that of steady-state heating, especially considering that the phonon-electron relaxation time has the finite value for the atomic structure of electron emitter. In an extreme case where the duration of heating is shorter than that of electron-phonon relaxation time, the electron and phonon conductions must be simulated separately and simultaneously. Therefore, the two-temperature proposed by Uimanov and coworkers [35] is employed in heat conduction (HC) module of FEcMD program. In the two-temperature heat conduction mechanism, the heat balance equations for phonons and electrons are given by Eqs. (22) and (23), respectively.

$$C_e(T_e)\frac{\partial T_e}{\partial t} = \nabla\left[\kappa_e\left(T_e, T_p\right)\nabla T_e\right] - G_{ep}\left(T_e - T_p\right) + \rho\left(T_p\right)j^2 \tag{22}$$

$$C_p(T_p)\frac{\partial T_p}{\partial t} = \nabla\left[\kappa_p\left(T_p\right)\nabla T_p\right] + G_{ep}\left(T_e - T_p\right) \tag{23}$$

Here, the subscripts 'e' and 'p' represent electrons and phonons, respectively. $C$, $T$ and $\kappa$ are the volumetric heat capacity, temperature, and thermal conductivity, respectively; $t$ is the heat conduction time; $\rho$ is the resistivity of the emitter; $j$ is current density; $G_{ep}$ is the electron-phonon coupling factor, which determines the energy exchange rate between electrons and phonons. Otherwise, the Nottingham heating of field emission occurs at the tip surface, and this part of the heat flux is given as a boundary condition for heat conduction in Eq. (24).

$$-\kappa_e \nabla T_e \,|_{\text{surface}} = q_N \tag{24}$$

Here, $q_N$ is the Nottingham heat flux, and which is given in Ref. [40]. In the field emission process, the Nottingham effect can either heat or cool the electron emitter according to the energy released or absorbed by the electrons passing through the potential barrier. The initial temperature of the emitter $T_{e,p}\,|_{t=0} = T_0$, and the temperature of bulk bottom $T_{e,p}\,|_{bottom} = T_0$.

In FEcMD program, Eqs. (22)-(24) are implemented in the open-source Deal.II library using the finite element mesh, enabling users to directly study the phonon and electron distributions and their evolution with time in ED simulation. Additionally, combining the two-temperature heat conduction mechanism with ED-MD simulation, the effects of electron temperature and phonon temperature on the structural evolution and electron emission characteristics can be investigated.

**3.4 FEcMD simulation setup**

In a typical FEcMD simulation, the micro-protrusion or nano-tip is divided into two segments, including the upper atomic structure and the lower coarse-grained region. The upper region of electron emitter is subjected to strong atomic structural deformations under high E-field and intense heating processes. It is expected that the upper region contributes most to the electron emission current. Meanwhile, the lower bottom is mainly employed to complete the full geometry of electron emitter, which also serves as thermal reservoir in the heat conduction calculation. In addition, the electron emission process of the lower coarse-grained segment of electron emitter is negligible. The geometry of the lower segment is fixed during the whole ED-MD-PIC simulation. The molecular dynamics simulation only applies to the upper region of the electron emitter. Nevertheless, the finite element mesh is generated for the entire electron emitter (atomistic upper part and coarse-grained extension). Practically, this technique greatly reduces the total number of atoms in the atomic structure of electron emitter, and which is critical to boost the overall computational efficiency of hybrid ED-MD simulations.

Regarding the dimensions of the simulation box, the height $H$ is usually set to $10h$, and $h$ refers to the total height of electron emitter. Meanwhile, the lateral dimensions of the box must be chosen carefully to accommodate a wide flat substrate to support the whole electron emitter, and which also acts as a thermal reservoir for heat transfer. The widths ($W$) of the box are set to $200 \sim 300 R_0$, and $R_0$ denotes the radius of electron emitter at the bottom that is attached to the flat substrate. This choice is made due to the fact that the space charge flux is negligible in the lateral directions for a sufficiently wide simulation box. The height of the atomistic region ($H_a$) of an electron emitter should be tested for different geometries of electron emitters. In the case of a conical nano-tip, $H_a$ is usually given as one half of the total height of emitter ($H_a = h/2$) [20, 27, 29, 30].

FEcMD software supports the ED-MD-PIC simulations for different geometries of nano-tip, as illustrated in Figure A1. The analytical expressions for different types of nano-tips are given in Table. A1. The procedures to create the nanotips are described in Appendix A. Notably, we provide the scripts in FEcMD software for users to create the initial atomistic structure of nano-emitter with different geometries and sizes. Additional scripts are interfaced with Deal II library and DBSCAN algorithms to build the coarse-grained extension of nano-emitter [21].

When conducting ED-MD-PIC simulations, it is crucial to run MD+AF, ED and HC modules in a hierarchical manner. Obviously, different time steps must be employed in PIC, ED, HC, and MD modules to obtain the space charge evolution, E-field distribution, temperature distribution and atomic structural charges in a multi-scale and multi-physics simulation self-consistently using FEcMD software. The smallest time step is needed to reliably track the motion of space charges and their inter-particle interactions in ED-PIC simulation. Based on our previous experiences and literature, the typical time step used in ED-PIC simulation should be smaller than 1 fs [20, 27, 29, 30]. It is important to keep the time step below this threshold value because the emitted electrons are rapidly accelerated by the high E-field and their velocities may quickly enter the relativistic regime. Using the small-time step ensures the accuracy of the algorithm to describe the trajectory of space charges and their transient spatial distribution. The latter information is critical to determine the electric potential and E-field with the presence of space charges by solving the Poisson equation in ED-PIC module. Therefore, the time-step to update the E-field on the finite element mesh is identical to that of PIC calculation. From the point view of time step hierarchy in the ED-MD-PIC simulation, molecular dynamics simulation is performed with a relatively large time step, compared to that of ED-PIC method. Since molecular dynamics simulation involves the change of atomic structures of nano-emitters, a reasonable time step is usually in the range from 1 fs to 5 fs. Using smaller time-step in MD simulation provides better resolution in particle dynamics with a higher computational cost. On the top of the time-step hierarchy, solving the heat balance equations for electron and phonon to obtain temperature evolutions in the electron emitter is conducted within the largest time step. Compared to the atomic structure or space charge evolution, heat transfer proceeds slowly on the timescale. Specifically, the typical time step adopted in the HC module is set to 30 ps ~ 50 ps. In ED-MD-PIC simulation, atomic velocities in MD simulation are adjusted on the same time interval to the time step of HC module.

All steps and procedures to conduct a typical ED-MD-PIC simulation are summarized as follows:

i. Using the scripts shown in Appendix A to build the nano-emitter. Then, performing the standard molecular dynamics simulation within MD+AF modules for 1 ps in the NVT ensemble to relax the atomic structure of nano-tip. The time step in the MD simulation is set to 4 fs ($\Delta t_{MD}$ = 4 fs, also See Figure 3).

ii. Executing the electrodynamics calculation in ED module, updating the electric field and field emission current for every 0.5 fs ($\Delta t_{PIC}$ = 0.5 fs, also see Figure 4).

iii. Employing the PIC technique in ED module to simulate the injection, collision, and motion of space charge, synchronously updating the electric field and emission current density at the same time step to that of ii.

iv. Counting surface atoms and calculating their net charges and forces (Coulomb force and Lorentz force) in AF module for every 4 fs to adjust the total atomic forces of atoms in nano-tip in MD module.

v. Calculating the Nottingham heat and emission current density and solving the coupled two-temperature heat transfer model in the HC module to obtain the electron and phonon temperatures of nano-tip with a constant time step of 40 ps ($\Delta t_{HC}$ = 40 ps, also see Figure 6). Using the obtained electron temperature at the surface of nano-tip to calculate the electron field emission property in ED module. The phonon temperature is returned to MD module, and velocities of atoms are rescaled to match the phonon temperature distribution profile obtained from HC module.

vi. Iteration restarts from stage ii, repeating until the user defined total duration of ED-MD-PIC simulation is achieved.

Examples and scripts are also distributed together with the source code of FEcMD software. All users are highly recommended to directly use or modify the provided inputs to build the nanotips and to conduct the ED-MD-PIC calculations.

**4. Validation and optimization of algorithms**

In this section, we would like to test and validate the implemented algorithms in the previous section. In addition, the use of different numerical recipes and their computational wall-times are discussed, and the optimization of different computational modules in FEcMD software are also demonstrated and addressed.

**4.1 Molecular dynamics simulation**

Here, we perform a series of molecular dynamics simulations using MD and AF modules in FEcMD software to validate the methodology, and also to estimate the effect of system size on the error in molecular dynamics. For such purposes, as shown in Figure 7, the conic Cu nano-tips with curvature radii of 1 nm, 5 nm, and 10 nm containing approximately $10^4$, $10^5$, and $10^6$ atoms are built, respectively. The standard molecular dynamics simulations are carried out for the three nanotips at 300 K using NVT ensemble and with a time step of 4 fs and a total duration of 1 ps. For all MD simulations, the lowest three atomic layers at the bottom of nanotips fixed to their initial atomic positions. Otherwise, we place each nano-tip in a large rectangular box. The periodic boundary conditions are enforced in the lateral directions for the simulation box. Meanwhile, the box is aperiodic in the vertical direction. The EAM potential of bulk Cu metal is adopted to describe interatomic interactions among atoms in nanotips. The initialization of the velocities of all free atoms in nanotip follows the Maxwell-Boltzmann distribution. Updating the atomic accelerations, velocities and coordinates is realized using the Leap-Frog time integration algorithm.

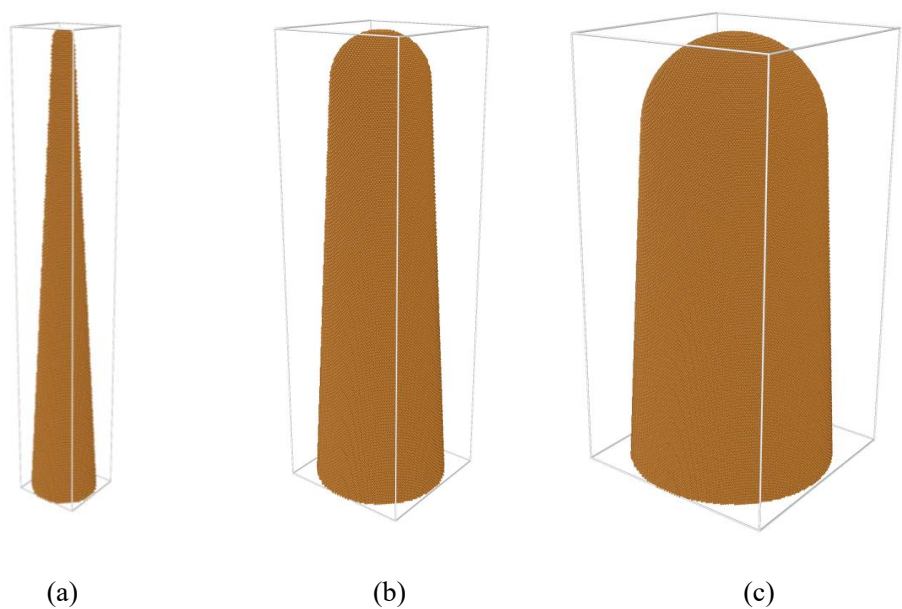

(a)　　　　　　　　　　(b)　　　　　　　　　　(c)

**Figure 7** Atomic structures of conic Cu nano-tips with a height $h$ = 50 nm, a half angle $\theta$ = 3°, and curvature radii of (a) $r$ = 1 nm (76992 atoms); (b) $r$ = 5 nm (487976 atoms); (c) $r$ = 10 nm (1489049 atoms).

For molecular dynamics simulation employing the classic inter-atomic potentials of various forms (pair potentials, EAM, and machine learning potentials), the most time-consuming and memory demanding step is the calculation of the interatomic forces by counting atomic pairs at different distances. In the FEcMD software, three methods including Neighbor-list, Cell division and All-pairs are implemented in MD module. In Figure 8, we demonstrate the scalability of the wall-time versus either the size of nanotips (total number of atoms) or the number of parallel threads using the three methods for 1 ps duration. As can be seen from Figure 8(a), the Neighbor-list method gives the best scalability for the total computational time versus total number of atoms in nanotips, i.e., the computational time is always the shortest among all three methods regardless of the structure size. Nevertheless, all three methods perform similarly for small nanotips containing atoms less than $10^3$. The All-pairs method exhibits relatively poor scalability for the total wall-time versus nanotip size. For large nanotip (> $10^4$ atoms), the Cell-division and Neighbor-list methods follow the same scaling behaviors as shown in Figure 8(a). Specifically, the All-pairs method counts all pairwise interactions in the system, resulting in a computational complexity of $O(N^2)$. This method is only suitable for small systems (<$10^3$ atoms), and as the number of atoms increases, the computation time increases significantly. For nanotips with ~$10^4$ atoms, the wall-time of the All-pairs method can be 50-150 times longer than the Neighbor-list or Cell division method, as shown in Figure 8(a). Both the Cell division and Neighbor-list methods divide the system into small cells or groups to calculate the interatomic interactions inside the cells and within neighboring cells. This strategy greatly reduces the total number of atoms in the calculation for both methods, leading to the $O(N)$ scaling performance for computational wall-time versus nanotip size. Regarding the computational scalability versus the number of parallel threads of MD module in FEcMD software, the tests are carried out using the Neighbor-list method, and the results are displayed in Figure 8(b) for three different nanotips. The multi-thread calculation is realized using the open-source OpenMPI.1.7.3 library. All MD tests are conducted on two Intel Xeon® CPUs (E5-2680 v4 @ 2.40GHz) with 56 cores. For the smallest nanotip containing 76992 atoms, the use of multi-thread computation brings little gain in the wall times, compared to that of single-thread computation, as shown in Figure 8(b). However, employing the multi-thread computation is indeed critical to reduce the computational wall-time for large nanotips. From Figure 8(b), it is obvious to see that the wall times do not scale linearly by increasing the number of CPU threads or cores in the task due to the increase of communication costs among processors. For the testing hardware employed here, the best wall-time saving is achieved using 8 CPU threads. Currently, FEcMD software does not support the multi-node parallel computing ability. The parallel scalability of MD module in FEcMD software on multi-node architectures is not performed.

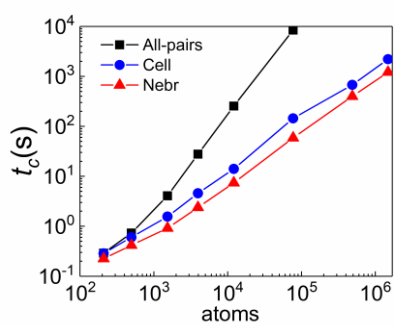
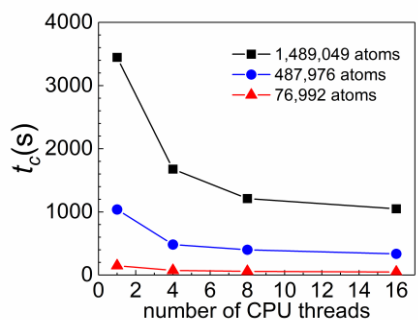

(a)　　　　　　　　　　　　　(b)

**Figure 8** Computational scalability of the algorithms (Neighbor-list, Cell division, and All-pairs) employed in MD module to calculate the interatomic forces for a copper nano-tip ($h$ = 50 nm, $r$ = 1 nm, $\theta$ = 3°) using NVT ensemble at T = 300 K for a total duration of 1 ps: (a) Wall-time consumption versus the total number of atoms in nano-tips; (b) Wall-time consumption versus number of parallel threads.

Next, we would like to demonstrate the convergence stability of MD module for performing the molecular dynamics simulation using NVT ensemble. For all tests, the Neighbor-list method is used for computing interatomic interactions, and 8 parallel threads on two CPUs (Intel Xeon® E5-2680 v4 @ 2.40GHz) are employed. The variations of the mean energies and their standard errors are shown in Figure 9 for 1 ps duration. For the temperature, all three nanotips contain sufficient number of atoms, ensuring that the simulated temperature quickly approaches that of target value (300 K) after 0.5 ps, as shown in Figure 9(a). Nevertheless, the standard deviation of the temperature is indeed found to be smaller for the large nanotips than that of the small one (76992 atoms) for the first several hundred fs in MD simulations. In Figures. 9(b) and 9(c), the variations of mean kinetic energy, potential energy, and total energy of the three Cu nanotips are illustrated for the MD duration of 1 ps. All quantities quickly converge to the values of the equilibrium state with the increasing of time steps. Notably, the calculated mean potential energy gets more negative when the size of nanotip is increased, as shown in Figure 9(b). Such a trend is mainly attributed to the increase of bulk Cu atoms in the large nanotip. As a result, the mean total energy per atom, as plotted in Figure.9(c), also becomes more negative going from r = 1nm to r = 10 nm. Finally, the overall scalability of computational wall-times within 1 ps MD task is displayed in Figure 9(d) using 8 parallel threads on two CPU processors for three nanotips. The good linear scalability for wall-time versus total number of atoms is observed up to $10^6$ atoms for the multi-thread task on two CPU processors.

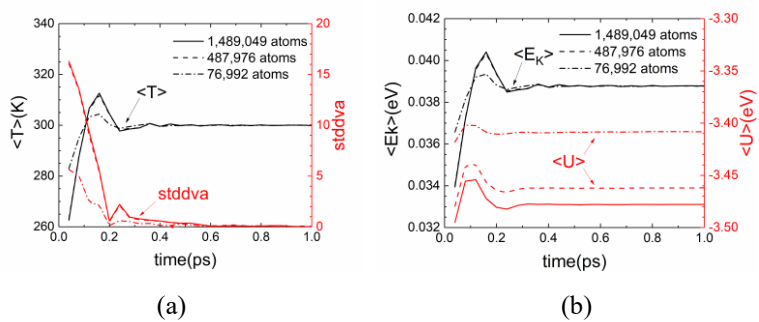

(a)  (b)

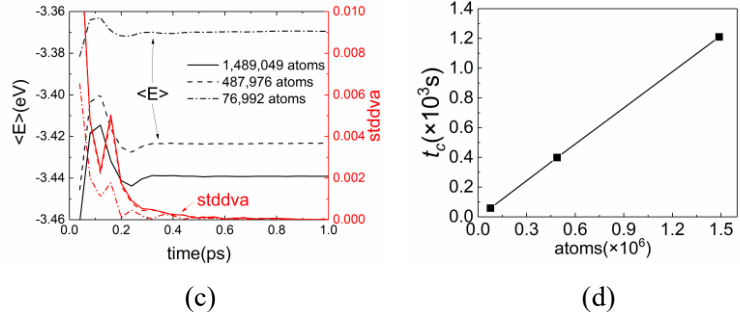

(c)　　　　　　　　　　　　　(d)

**Figure 9** Convergence tests for MD module in FEcMD software and its wall-time scalability: (a) Variations of the temperature and its standard deviation over time; (b) Variations of the mean kinetic energy and potential energy, and their standard deviations with accumulated time-step; (c): Variation of the mean total energy per atom versus the total time-step and its deviation; (d): Wall-time scalability with the nanotip size.

**4.2 Electrodynamics module**

To test the accuracy and convergence of algorithms implemented in ED module for electrodynamics, we perform the benchmark tests for a three-dimensional model as shown in Figure 10. The model represents a metal hemisphere on a flat planar surface, and which has a well-known analytical solution for solving the Laplace equation in the surrounding space [60]. In the benchmark tests, we numerically solve Laplace equation for the model to obtain the E-field distribution using both finite difference method (FDM) and finite element method (FEM). Then, the maximum electric field and field enhancement factor at the top of the hemisphere are also calculated. As can be seen from Figure 10, the simulation is conducted in a box with finite simulation volume. Nevertheless, when the simulation volume is sufficiently large, the calculated E-field distribution around the hemisphere, and the field enhancement factor at the top are converged to those of analytical results. Regarding the dimensions of simulation box shown in Figure 10, the heigh of the box in the vertical direction is given as $d_\mathrm{p} = 6R$, and the widths of the box in the horizontal directions are set as $d_\mathrm{h} = 10R$, and $R$ denotes the radius of hemisphere. Here, we describe the numerical algorithms for solving Poisson equation in ED module and setting the charge density $\rho = 0$ gives the results for Laplace equation.

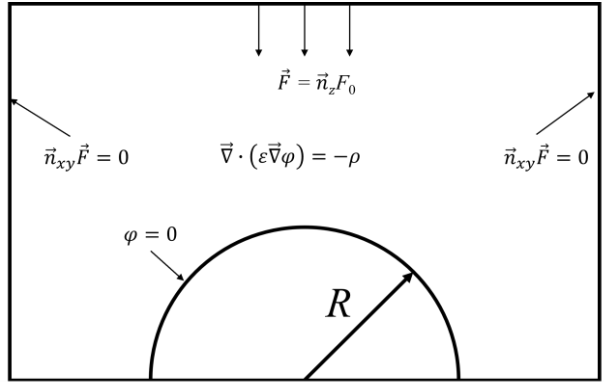

**Figure 10** Schematic diagram of the benchmark model for electric field calculation using ED module in FEcMD software. The model represents a hemisphere on the flat planar substrate in the vacuum without free charge. The boundary conditions employed to solve the Laplace equation are indicated ($\rho = 0$).

The finite-difference method (FDM) involves discretizing the Poisson equation on the uniform mesh grid for the simulation volume, as shown in Figure 10. The Poisson equation (or Laplace equation by setting $\rho = 0$) is solved using Eq. (25).

$$\frac{\varphi_{i-1,j,k} + \varphi_{i+1,j,k} + \varphi_{i,j-1,k} + \varphi_{i,j+1,k} + \varphi_{i,j,k-1} + \varphi_{i,j,k+1} - 6\varphi_{i,j,k}}{d^2} = -\frac{\rho}{\varepsilon} \quad (25)$$

Here, subscripts $i$, $j$, and $k$ represent the IDs of FDM mesh points in the $x$, $y$, and $z$ directions, respectively, and $d$ denotes the spacing between the two grid points on the uniform mesh. The discretization of the simulation volume in the box using the uniform grid is illustrated in Figure 11 (a).

The discretization of Poisson equation in the FEM solver is more complicated, and the derivations are presented in Appendix F. To build the FEM mesh grids, FEcMD software employs the self-adaptive mesh generation technique, providing multi-scale mesh quality resolution. Specifically, we divide the box into three regions: the top of the hemisphere where the electric field is highly non-uniform, which requires the highest level of mesh refinement to accurately capture the field variations; a coarser mesh is used in the middle and lower parts of the hemisphere; the E-field is relatively uniform in the upper vacuum region, and an even coarser mesh can be employed. The specific definitions of mesh division and coarsening factors can be found in [21]. For the numerical implementation, the DUBSCAN algorithm is adopted to divide the mesh nodes into different simulation regions [55]. After dividing the mesh nodes, we use the open-source software package Tetgen for mesh generation based on Delaunay triangulation in three-dimensional space [61]. To improve computational efficiency and accuracy, each tetrahedron is further divided into four hexahedrons, and the specific division method can be found in Ref.[21]. In Figure 11 (b), the FEM mesh is shown for the benchmark simulation model. The dense mesh is created at the interface between the metal hemisphere and vacuum. The coarse mesh is employed in the vacuum region that is located far away from the hemisphere.

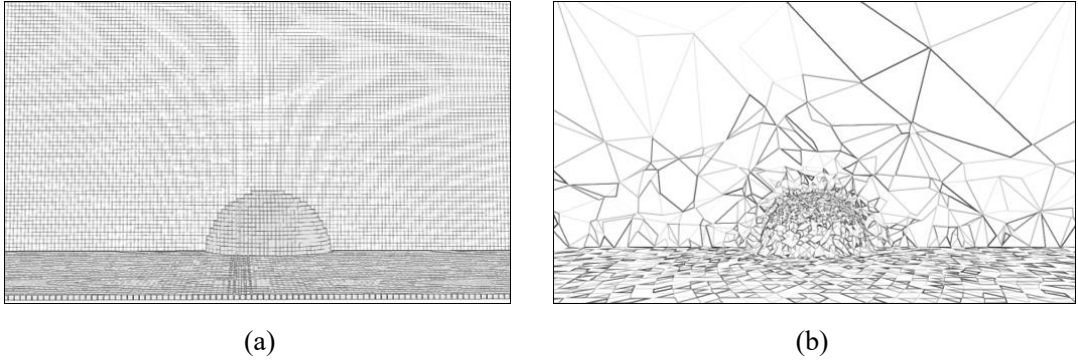

(a)  (b)

**Figure 11** Mesh grids used to solve the Laplace equation in the benchmark tests using FDM and FEM: (a): Uniform mesh grid; (b): the finite element mesh grids with multi-scale resolution quality. The height of the simulation box is $d_p = 6R$, and the horizontal width of the box is $d_h = 10R$.

The E-field distribution of the benchmark model is illustrated in Figure 12 for the numerical results and analytical solution on the surface of metal hemisphere. Overall, the numerical calculations using FDM and FEM agree with that of analytical solution. The FEM provides better resolution of the E-field distribution on the top of metal hemisphere due to the use of the finer mesh than that of FDM.

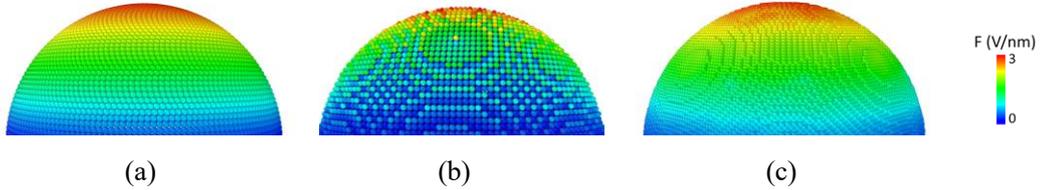

(a)  (b)  (c)

**Figure 12** Electric field distribution on the surface of the hemisphere (a) Analytical solution; (b) Finite difference method (FDM), with a minimum grid volume of 0.006 nm$^3$; (c) Finite element method (FEM), with a minimum grid volume of $1.18 \times 10^{-4}$ nm$^3$. The radius of the sphere $R = 3$ nm, and the externally applied long-range electric field $F_0 = 1$ V/nm.

The accuracy of the numerical results is displayed in Figure 13 for both FDM and FEM. Here, we calculate the maximum E-field value at the top of metal hemisphere and compare the numerical results with that of analytical solution. Obviously, the computational errors for both methods are decreased by increasing the mesh density. However, FEM solver shows higher accuracy and lower computational wall-time than those of FDM solver. To reach the same level of accuracy using FDM solver, the very fine mesh must be used in the calculation, leading to a significant increase in wall-time, compared to that of FEM solver using the similar mesh density. Thus, the FEM solver exhibits the good scalability of computational wall-time versus the mesh density.

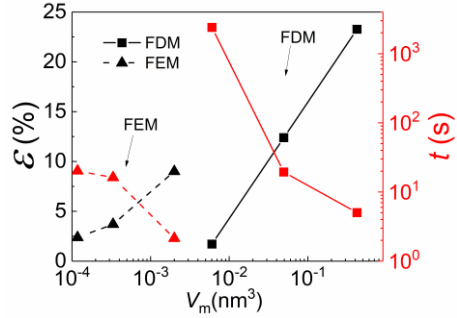

**Figure 13** Computational wall-times and errors of FDM and FEM solvers using different mesh densities for solving Laplace equation to obtain the maximum electric field on the top of a metal hemisphere. All calculations are performed using 8 parallel threads on two CPUs (Intel Xeon® E5-2680 v4 @ 2.40GHz).

To further improve the computational efficiency of FDM solvers in ED module, we employ three iterative methods, including Jacobi, Gauss-Seidel, and Over-relaxation solvers. From Figure 14(a), it is observed that Gauss-Seidel and Over-relaxation ($\omega = 1.2$) methods decrease the wall-time by 6-9 times, compared to that of Jacobi solver. At the same time, three iterative methods have almost no impact on computational accuracy.

Otherwise, Poisson or Laplace equation is transformed into solving large sparse matrices in FEM method. Three algorithms are implemented in FEcMD software, including conjugate gradient (CG), generalized minimal residual (GMRES) and biconjugate gradient stabilized (BICGSTAB) methods. Although CG solver is easy to implement and is also widely used in FEM method, it is only suitable for symmetric positively defined matrices. Notably, GMRES and BICGSTAB solvers are more suitable for general matrices in FEM method. GMRES uses the generalized minimal residual method to manage the high demand of storage space and memory through either clearing or truncating the previous vector sequences during the iteration. Otherwise, BICGSTAB generates two sets of mutually orthogonal vectors similar to CG for solving non-symmetric invertible matrices. It is similar to the BICG method but mitigates the irregular convergence issue in the latter method, requiring additional computational efforts. From Figure 14(b), it can be seen that all three methods are suitable for iterative solution of the Poisson equation when the employed mesh densities are the same. Furthermore, the computational efficiency and accuracy are also similar for the three solvers in ED module. In general, the FEM method is highly recommended to solve the Poisson or Laplace equation in FEcMD software. Compared to FDM solvers, FEM solvers provide the lower computational cost and higher numerical accuracy.

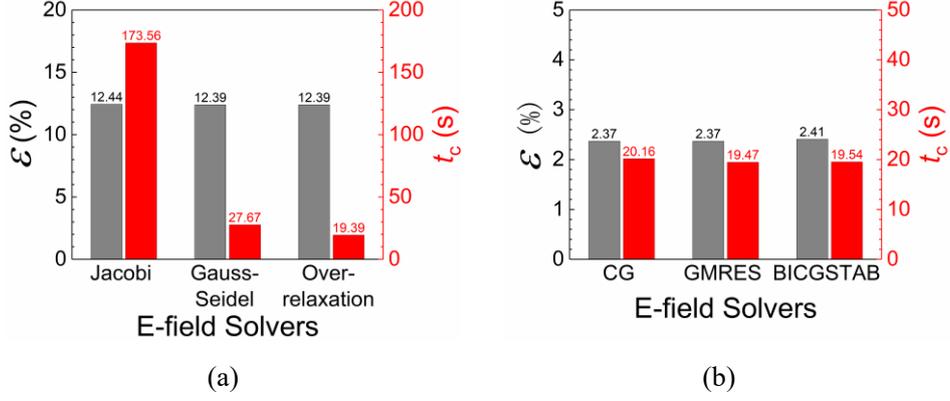

(a)                                                (b)

**Figure 14** Errors and computational wall-times of FDM and FEM methods employing different numerical solvers for benchmark tests.: (a) FDM with a minimum grid volume of 0.049 nm$^3$; (b) FEM with a minimum grid volume of $1.18 \times 10^{-4}$ nm$^3$. All calculations are performed using 8 parallel threads on two CPUs (Intel Xeon® E5-2680 v4 @ 2.40GHz).

**4.3 Heat conduction module**

In FEcMD software, the heat conduction module solves the heat balance equations for phonon and electron subsystems using the FEM solvers. We implement the two-temperature heat transfer model in open-source Deal II library. The discretization of Eqs. (22) and (23) within FEM solvers is addressed in Appendix G, resulting in the two-temperature heat balance equations in the matrix forms of Eqs. (26) and (27).

$$\mathbf{C}_e \cdot \frac{\partial \mathbf{T}_e}{\partial t} + \mathbf{K}_e \cdot \mathbf{T}_e = \mathbf{f}_e \tag{26}$$

$$\mathbf{C}_p \cdot \frac{\partial \mathbf{T}_p}{\partial t} + \mathbf{K}_p \cdot \mathbf{T}_p = \mathbf{f}_p \tag{27}$$

We adopt the forward difference method to discretize the time component in Eq. (28) and introducing the relaxation factor $\Theta$ in Eq. (29). Three values are available for $\Theta$, i.e., Euler ($\Theta = 0$), Crank-Nicolson ($\Theta = 0.5$) and the implicit Euler ($\Theta = 1.0$) scheme, respectively. For the hybrid MD-HC simulation, the implicit Euler method gives better numerical convergence performance than other schemes. Therefore, the implicit Euler scheme has been used for all benchmark tests presented below.

$$\frac{\partial \mathbf{T}}{\partial t} = \frac{\mathbf{T}^{n+1} - \mathbf{T}^n}{\Delta t} \tag{28}$$

$$\mathbf{T}^{n+\Theta} = \Theta \mathbf{T}^{n+1} + (1-\Theta)\mathbf{T}^n \tag{29}$$

Coupling HC module with ED-PIC method is straightforward, because both methods use the same set of finite element mesh to solve the differential equations within the open-source Deal II library. In the ED-HC calculation, the electron temperature needs to be updated and passed on to the ED module for field emission calculation after each iteration with a time step of $\Delta t_{HC}$. On the other hand, the phonon temperature distribution is used to adjust velocities of atoms in nanotip within the same time interval. To achieve such purpose, the nanotip is divided into a finite number of thin slabs in the axial direction with a thickness of $\Delta z$ for each slice. Then, the finite element cells and atoms in the interior of nanotip are grouped based on their z coordinates. Solving the heat balance equations for two-temperature model gives the phonon temperature distribution as a function of tip height for each slab using FEM. This temperature profile defines the macroscopic temperature of the nanotip in the axial direction. Then, atomic velocities are rescaled for all atoms residing in the same slab (See Eq. (9)). For the very large electron emitters, the temperature distributions in both radial direction and axial direction are non-uniform. In this case, each thin slab in the axial direction of the electron emitter is further divided into multiple small cells in the radial direction. The atoms and finite element mesh grids are grouped both in radial and axial directions, enabling the rescaling of atomic velocities in each small cell within the same thin slab. For the benchmark tests presented below, we evaluate the accuracy and performance of both cell division schemes for a conical Cu nanotip using FEcMD software.

For the benchmark tests, we build a conical Cu nanotip containing 76992 atoms (h = 50 nm, $\theta$ = 3° and r = 1 nm). To initiate the resistive heating process (Joule heat) due to electron field emission mechanism, an external electric field of 250 MV/m is applied in all calculations using ED-MD-PIC scheme. The iteration time steps for MD, ED-PIC and HC modules are set as $\triangle t_{MD}$ = 4 fs, $\triangle t_{PIC}$ = 0.5 fs, $\triangle t_{HC}$ = 40 fs, respectively. Therefore, rescaling of atomic velocities is performed for atomistic simulation in every 40 fs. Different slicing grids for nanotips are tested including 3×3×100, 1×1×20 and 1×1×100. The temperature and size dependencies of electrical resistivity of Cu nanotip are obtained from previous works [23]. The initial temperature of the nanotip is set to 300 K.

The phonon temperature evolution and the effects of rescaling process on the atomic kinetic energies are illustrated in Figure 15. It is found that phonon temperature of the nano-tip rapidly increases with the increasing of time steps, forming a temperature gradient of approximately 1000 K in the z direction of the Cu nano-tip within the first 10 ps duration (See Figure 15(a)). The apex of nanotip is subjected to stronger resistive heating than the lower part. After rescaling velocities, atomic kinetic energy shows a noticeable increase in the upper part of nanotip (See Figure 15(b)). Notably, rescaled atomic velocities still follow the Maxwell-Boltzmann distribution.

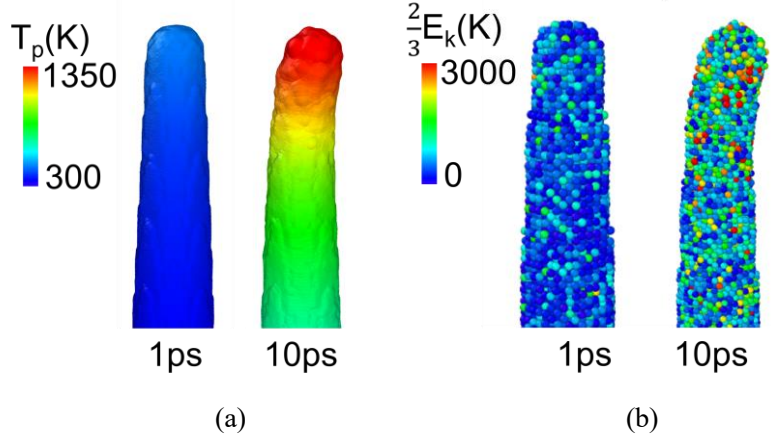

**Figure 15** Phonon temperature and atomic velocities rescaling in a hybrid ED-MD simulation for Cu nanotip containing 76992 atoms (h = 50 nm, $\theta$ = 3° and r = 1 nm): (a) Phonon temperature evolution; (b) Atomic kinetic energy before and after rescaling step. The cell dividing grid for temperature control is 3×3×100. The simulation is performed using 8 parallel threads on two CPUs (Intel Xeon® E5-2680 v4 @ 2.40GHz).

To quantify the precision and efficiency of temperature rescaling algorithm, we calculate the mean error (ME) in local temperatures between atomistic simulation and FEM results for divided cells, as given by Eq. (30).

$$\mathrm{ME} = \frac{1}{N_p}\sum_{j=1}^{N_p} \frac{T_p - \frac{2}{3}\sum_{i=1}^{N} E_{ki}\Big/N}{T_p} \tag{30}$$

Here, $N$ and $N_p$ represent the total number of atoms and total number of finite element mesh points in the same divided cell, respectively. First, we test the influence of the divided cell size on the ME of temperature, and the results are shown in Figure 16(a) as a function of simulation time. When the temperature distribution in the system tends to be uniform, such as when the initial temperature is close to 300 K throughout the whole nanotip at the first several ps, using a dense dividing grid can result in small cell volume. As a result, each cell may contain few atoms (< 100). Thus, statistical error is signified in calculating the average atomic temperature. As can be seen from Figure 16(a), the ME of 3×3×100 dividing grid is ~1% higher than those of 1×1×20 and 1×1×100 grids in the first 2 ps. As the temperature gradient in the Cu nanotip gradually develops with the increasing of MD duration, increasing the grid density in the axial direction is needed to reduce the ME that is caused by temperature non-uniformity. For example, the ME of 1×1×100 dividing grid is halved to that of 1×1×20 grid at 10 ps. Furthermore, the computational wall-time versus total number of dividing cells is displayed in Figure 16(b). The total number of atoms contained in each cell is reduced by increasing the number of cells. Meanwhile, the wall-time slightly increases with the number of cells. For example, using 900 cells only takes 6.6% more computation wall-time than that of using 100 cells. In summary, cell dividing in both axial and radial directions provides better numerical accuracy for large nanotip than that of axial cell only. Otherwise, the axial cell is sufficient for small nanotip.

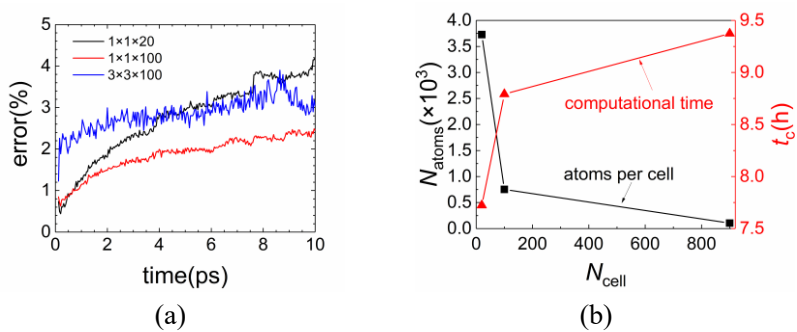

**Figure 16** Accuracy and computational efficiency of phonon temperature rescaling algorithm implemented in HC module: (a) Variations of the mean error (ME) with simulation time for different cell dividing grids; (b) Variations of the number of atoms per cell and total computation wall-time versus the number of cells. All calculations are conducted using 8 parallel threads on two CPUs (Intel Xeon® E5-2680 v4 @ 2.40GHz) processors for Cu nanotip containing 76992 atoms (h = 50 nm, $\theta$ = 3° and r = 1 nm).

**4.4 Computational wall-time breakdown of FEcMD simulation**

Here, we breakdown the total computational cost for completing two ED-MD-PIC simulation task within 40 fs. This duration represents the shortest MD time interval that is required by the HC module to update the temperature profiles in nanotip. The Cu nanotips illustrated in Figure 7 are used for the tests. All numerical calculations are conducted on two CPU (Intel Xeon® E5-2680 v4 @ 2.40GHz) processors using different parallel thread numbers. The breakdown of computational cost is achieved by decomposing the total computational wall-time into different computational modules, including MD+AF, ED-PIC, HC, and Mesh updating/generation. The results are plotted in Figure 17 for three Cu nanotips. It is interesting to see that ED module represents the most wall-time demanding step in ED-MD-PIC simulation using FEcMD software for the small nanotips, i.e., >60% of total wall-time (See Figure 17(a)).

Meanwhile, the wall-time needed to update the finite element mesh accounts 15%~30% of total wall-time for the smallest Cu nanotip (76992 atoms), and that value quickly goes up to 40%~60% for larger nanotips (See Figures 17(b) and 17(c)). Therefore, the mesh generation denotes the most time-consuming step in an ED-MD-PIC simulation for large nanotip in the current workflow of FEcMD software. Notably, wall-time consuming of MD module in ED-MD-PIC simulation is negligible for small nanotip, compared to wall-times of other steps. On the other hand, HC module refers to the least time-consuming step for large nanotips. From Figure 17, we conclude that MD+AF and HC modules show very promising scalability for the wall-time with respect to the number of parallel threads and the total number of atoms. For those two steps, no significant increase of wall-times is observed with the increasing the total number of atoms.

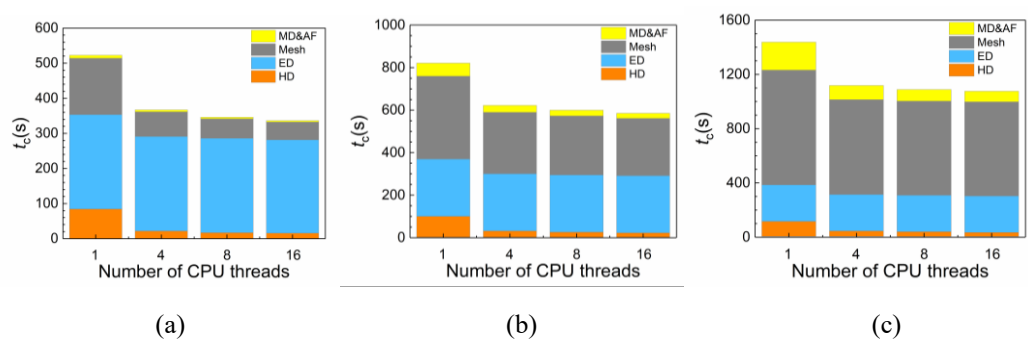

(a)  (b)  (c)

**Figure 17** Breakdown of computational costs for completing two ED-MD-PIC simulation for 40 fs in FEcMD software for Cu nanotips: (a) 76992 atoms (h = 50 nm, $\theta$ = 3° and r = 1 nm); (b) 487976 atoms (h = 50 nm, $\theta$ = 3° and r = 5 nm); (c) 1489049 atoms (h = 50 nm, $\theta$ = 3° and r = 10 nm). All calculations are conducted on two CPUs (Intel Xeon® E5-2680 v4 @ 2.40GHz) processors using different parallel threads.

## 5. Applications

Here, we demonstrate the capabilities of FEcMD software package by performing the numerical simulations for several different applications: 1. Applying the two-temperature heat conduction model for ED-MD simulation of Cu nanotip under RF electric field; 2. Electron emission and atomic structural evolution of Cu nanotip with and without the exchange-correlation potentials; 3. The formation of nano-protrusions on planar electrodes made of Cu and W-Mo alloys; 4. ED-MD simulations for Cu and W-Mo alloys using MTPs; 5. Electrodynamics of different geometries for electron emitters.

**5.1 Heat conduction of Cu nanotip under two-temperature model**

Here, we employ the two-temperature thermal conduction model implemented in ED module to illustrate the decoupling of phonon and electron temperatures in the Cu micro-protrusion under the RF electric field. The RF electric field is defined by the angular frequency of 10 GHz and the amplitude of 300 MV/m. For Cu micro-protrusion is approximated by a conical stand terminated with a hemispherical cap. The total height and the radius of the hemispherical cap of Cu micro-protrusion are set to 2 μm ($h_0$) and 22.5 nm ($r_0$), respectively. Additionally, the conical tip has a half-aperture angle ($\theta_0$) of 3°. The thermophysical properties of electron and phonon of Cu are also needed to perform the two-temperature heat conduction simulation (See Eqs. (22)-(24)). All parameters employed in our simulations can be found in Ref [35]. Notably, the lattice thermal conductivity of Cu is set to zero in the current study mainly because the phonon heat conduction is significantly smaller than that of electron. Otherwise, the electric conductivity of Cu has also been corrected by the temperature and nano-size effects [62-64]. Since electron thermal conductivity of Cu nanotip is obtained from the Wiedemann-Franz law with the Lorentz number $L = 2.0 \times 10^{-8}$ W Ω K$^{-2}$, the electron thermal conductivity has the dependences on both the size and temperature of nano-tip [65]. Finally, the electron-phonon coupling constant $G_{ep} = 4 \times 10^{17}$ W m$^{-3}$ K$^{-1}$ [66-68]. The geometry of Cu micro-protrusion is rigid, and which is fixed to the initial shape during the ED simulation. The initial temperature of the Cu nanotip is set to 300 K and the total duration of the ED simulation lasts for 2 ns.

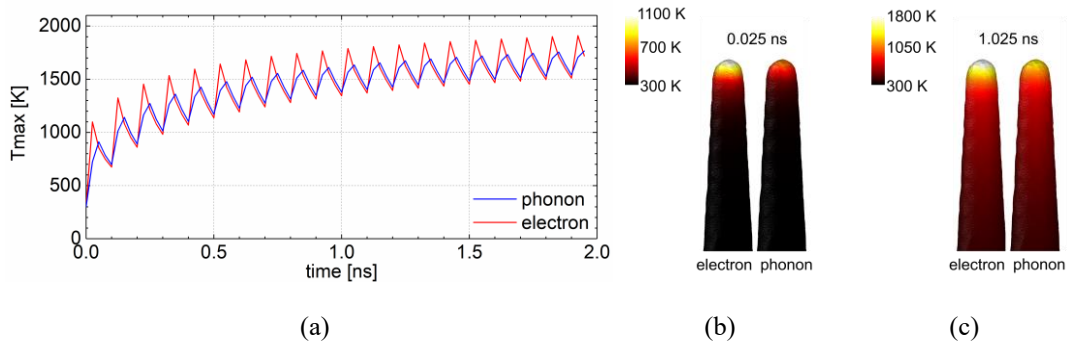

(a)            (b)            (c)

**Figure 18** The maximum temperature of electrons and phonons versus time at the apex of a conical Cu ($r_m$ = 22.5 nm, $h_m$ = 2 μm) micro-protrusion, respectively, under a radiofrequency electric field (F = 300 MV/m, 10 GHz). (b) and (c) shows the corresponding electron and phonon temperature distributions of micro-protrusion at 0.025 ns and 2.025 ns, respectively.

The variations of phonon and electron temperatures versus time profiles are illustrated in Figure 18(a). The results clearly demonstrate the decoupling of electron and phonon temperatures in Cu micro-protrusion under RF electric field. More specifically, we observed the electron temperature is always higher than that of phonon during the first half-cycle (50 ps). Certainly, the electron emission process and the associated heating mechanisms (Joule and Nottingham heats) at the apex region of Cu micro-protrusion must coincide with the periodicity of RF electric field (10 GHz). The duration of heating process now is comparable to that of electron-phonon relaxation time, and the decoupling of electron and phonon temperature evolutions are anticipated. It is also worth mentioning that the difference between phonon and electron temperatures inside the Cu micro-protrusion does not diminish in the entire simulation duration under the RF electric field. In Figures. 18(b) and 18(c), we also display the phonon and electron temperature distributions in the interior of Cu micro-protrusion at different simulation stages. Obviously, the electron and phonon temperatures show large difference at the apex region of Cu micro-protrusion in the beginning of simulation (0.025 ns). Meanwhile, the temperature distributions of electron and phonon subsystems are gradually synchronized at the late stage of ED simulation using the two-temperature model, indicating that the one-temperature model is indeed a valid approximation when either the electric field has the weak dependence on time or the on-set electric pre-breakdown time is sufficiently long, compared to that of electron-phonon relaxation.

To further understand the effects of E-field frequency on the temperature evolution in the two-temperature model, additional ED-MD-PIC simulations are carried out for the same Cu nanotip under a shorter pulse duration (100 GHz) within 0.5 ns. The maximum temperature profiles of electron and phonon subsystems are shown in Figure 19 for both 10 GHz and 100 GHz pulse durations. It is found that the electron temperature profile fluctuates strongly at the same rate as that of pulse duration. Otherwise, fluctuation in phonon temperature is greatly suppressed by increasing the pulse frequency from 10 GHz to 100 GHz. In the two-temperature model, heat dissipates through electron and phonon conductions in nanotip. The electron channel could remove the resistive heating more effectively and rapidly than that of phonon conduction for the Cu nanotip. Therefore, heating and cooling of nanotip due to the electron emission closely follow the pulse duration. On the other hand, phonon conduction proceeds rather slowly for heavy element, indicating that the maximum phonon temperature may not rise as fast as that of electron within a pulse period. Notably, the electron-phonon coupling constant ($G_{e-p}$) also significantly affects the rise of the maximum phonon temperature. By increasing the $G_{e-p}$ value (See Eqs. (22) and (23)), the heat exchange rate between phonon and electron subsystems is enhanced. As a result, the maximum phonon temperature could rise substantially with the increasing of $G_{e-p}$ in two-temperature model [35].

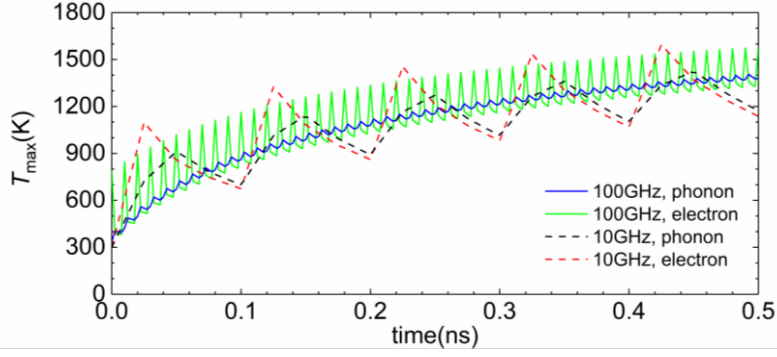

**Figure 19** The maximum temperature profiles of electron and phonon subsystems in two-temperature model under different E-field pulse durations (100 GHz and 10 GHz) using FEcMD software for Cu nanotip. The amplitude of E-field is fixed to 300 MV/m.

In our previous works, we investigated the electric pred-breakdown characteristics of metal nanotips, and concluded that the typical pre-breakdown time for small metal nanotips ($R_0 <$ 10 nm) could be on the scale of tens of ps. The decoupling of phonon and electron heat conduction mechanisms should be strong for nano-tips even under a static E-field. Motivated by the above consideration, additional ED-MD simulations are also conducted for Cu nano-tip ($r_0$ = 1 nm, $h_0$ = 100 nm and $\theta_0$ = 3°) with $E$ = 300 MV/m. The total duration of the ED-MD simulation is set to 40 ps because the electric pre-breakdown time for a nano-tip with this size is predicted to be on the same time scale. The temperature evolution profiles with and without using the two-temperature heat conduction model in ED-MD simulations are depicted in Figures. 20(a) and 20(b). Meanwhile, the temperature evolution in the Cu nano-tip is shown in Figure 20(c). Specifically, the maximum temperature versus time profiles at the apex region of Cu nanotip are shown in Figure 20(a), and average temperature versus time profiles are plotted in Figure 20(b). In both Figures. 20(a) and 20(b), we observe the very rapid increasing of temperature for electron subsystem initially in the ED-MD simulation. For example, the highest temperature at the apex of nano-tip can reach 2500 K instantly after the electron emission started. On the other hand, the phonon temperature also increases quickly with simulation time, but less prominent than that of electron subsystem. After performing ED-MD simulations for 10 ps, both the maximum and mean temperatures start to decrease. Notably, the electron temperature decreases more rapidly than that of phonon temperature. Interestingly, electron temperature eventually converges to that of phonon when the ED-MD duration is longer than 30 ps, as shown in Figures. 20(a) and 20(b). Finally, we show the 3-D temperature distributions of electron and phonon subsystems in Figure 20(c) for Cu nano-tip at different time step sizes. Clearly, the electronic heat conduction dominates the heating process at the beginning of ED-MD simulation, and two subsystems show very different temperature evolution profiles. Nevertheless, the difference in temperature profiles of electron and phonon subsystems is diminished after running ED-MD simulation for 20 ps. In conclusion, the implementation of the two-temperature heat conduction model in FEcMD program is critical to understand the heating process for micro- or nano-protrusion under both RF and static E-fields.

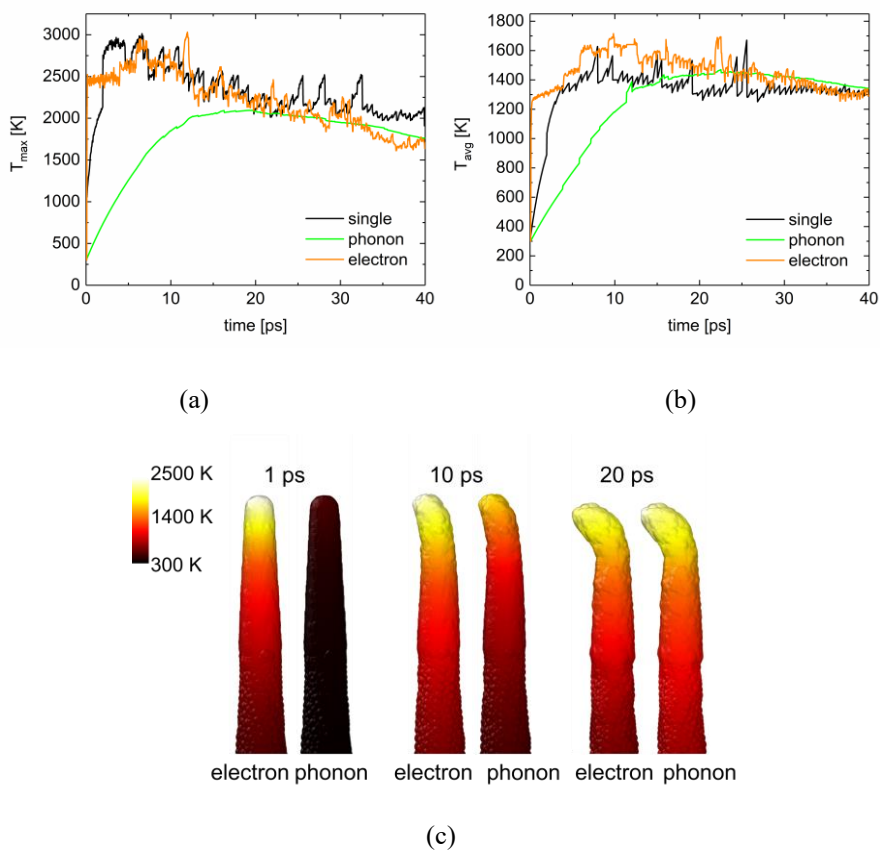

**Figure 20** The evolution of the maximum temperature (a) and the average temperature (b) of a conical Cu nano-tip ($r_0$ = 1 nm, $h_0$ = 100 nm and $\theta_0$ = 3°) versus time under a static electric field (300 MV/m) using both the single-temperature and two-temperature models, respectively; (c) electron and phonon temperature distributions of Cu nano-tip at 1 ps, 10 ps and 20 ps, respectively.

**5.2 Structural evolution of Cu nanotip with space charge effects**

In this section, we would like to address the effects of inclusion of space charge fields (space charge potential and exchange-correlation potential) in the WKBJ model on the electron emission properties and atomic structural evolution of Cu nano-tips using FEcMD software. It is worth noting that the inclusion of space charge exchange-correlation effects and space charge potential plays the critical role to understand the electron field emission characteristics of nanogap [59].

The Cu nanotip employed here for ED-MD simulations has the same geometry as that of section 4. 1. The exchange-correlation effects are treated at the level of local density approximation (LDA) using the Slater exchange functional [69] and Perdew-Zunger correlation functional [70], respectively. The ED-MD simulations are also performed for the same Cu nano-tip without considering the space charge fields. For all ED-MD simulations, the static E-field is employed, and the value is set to 320 MV/m.

In Figure 21(a), the space charge distribution is displayed for Cu nano-tip ($h_0$ = 100 nm, $r_0$ = 1nm and $\theta_0$ = 3°) after running the simulation for 1 ps. The space charge density is mainly distributed in front of nano-tip, and which is also highly concentrated at the apex region of nano-tip. The results are consistent with the fact that the local electron emission is expected to be strong at the tip of conical nano-emitter due to the high concentration of electric field strength. In Figures. 21(b) and 21(c), the local emission current density distribution and exchange-correlation potential are shown near the tip surface. Obviously, there is a strong correlation between the electron emission intensity and quantum many-body effects. To be more specific, electron emission is enhanced by the presence of exchange-correlation potential. Although the enhanced electron emission process could increase the local space charge density in front of Cu nano-tip, the space charge potential is also increased. The exchange-correlation effects of space charges are negative, and which can lower the electron emission energy barrier height. Meanwhile, the space charge potential among emitted electrons is always positive, which tends to increase the emission energy barrier. Nevertheless, the exchange-correlation effects could overwhelm the space charge potential for nano-gap with high emission current density [39, 71]. As shown in Figure 21(d), the overall electron emission barrier is greatly lowered after considering the space charge fields (space charge potential and exchange-correlation effects) in front of nanotip surface, compared to the ED-MD simulation without considering the space charge fields. Notably, the image charge potential is calculated using its classic form as given by Eq. (20) to plot Figure 21(d). The image charge potential represents the long-range asymptotic form of exchange-correlation effects between the emitted electrons and those in the electrode. Therefore, the image charge potential can lower the electron emission barrier height. Obviously, the classic image charge potential has the singularity at the interface between the electrode and vacuum when $x$ = 0. As a result, the electron tunneling energy profiles presented in Figure 21(d) is too negative and unphysical for small $x$. The singularity in image charge potential can be eliminated using the more advanced theoretical models such as free electron Thomas-Fermi approximation (TFA) with the random phase approximation [58]. Previously, we have already employed such a methodology to study the field emission property of nanogaps within a simplified one-dimension (1-D) model. Here, the method is further expanded to the full three-dimensional (3-D) nanotips in FEcMD software. The electron emission barrier height profiles obtained using the classic image charge model and advanced TFA are compared with each other in Figure 21(d) with and without space charge effects. It is easy to see that the TFA model eliminates the singularity in energy barrier profiles at $x = 0$ regardless of the space charge effects (exchange-correlation potential and space charge potential). The influence of local radius of curvature (RoC) for non-planar electron emitter on the calculated emission energy barrier profile and local electrostatic potential are displayed in Figure 21(e). Bending of electrostatic potential is clearly seen in Figure 21(e). For sharp electron emitters, local electrostatic potential is significantly overestimated by assuming the planar geometry at the large distance, compared to that of numerically interpolated value. As a result, the obtained electron emission energy barrier profile for planar case is above that of non-planar geometry. However, the electrostatic potential profiles of planar and non-planar geometries converge at the very small distance (< 0.1 nm) norm to the emission surface. Overall, the local RoC must be considered for sharp electron emitters in calculating the electrostatic potential and field emission properties.

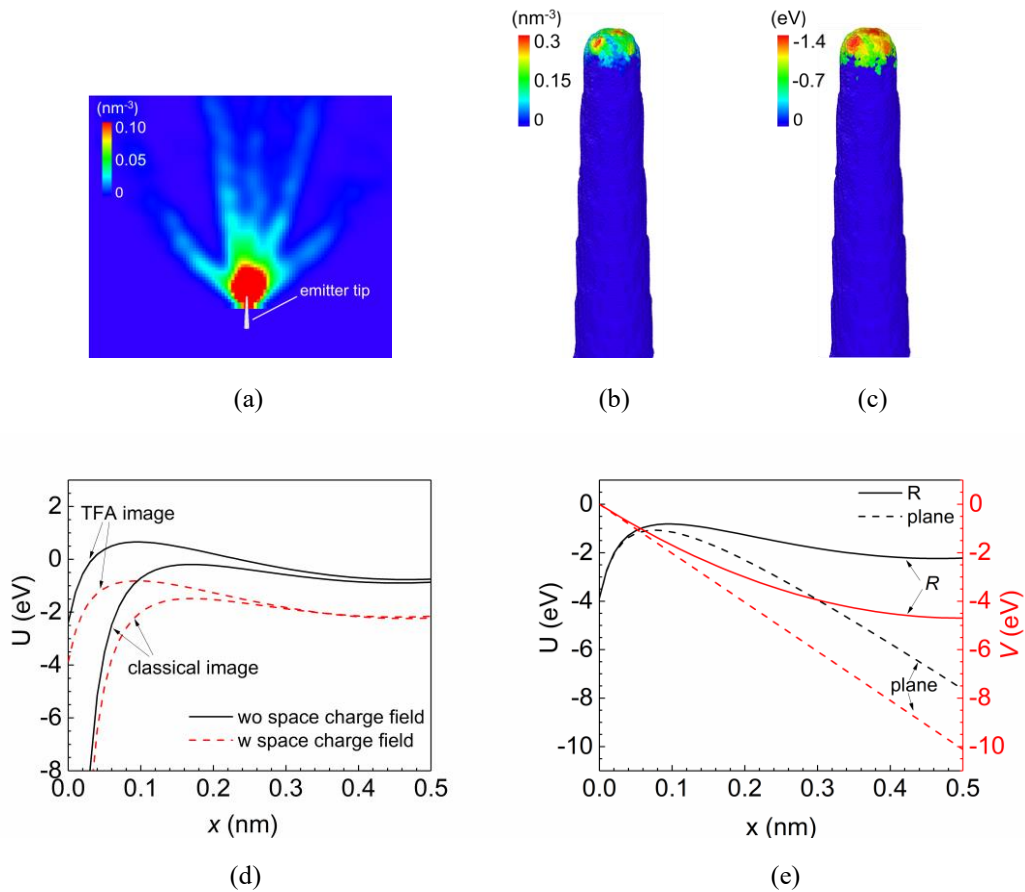

(a)　　　　　　　　　(b)　　　　　　　　　(c)

(d)　　　　　　　　　　　　　　(e)

**Figure 21** The space-charge density distribution cross-section of Cu nano-emitter in front of tip ($h_0$ = 100 nm, $r_0$ = 1nm and $\theta_0$ = 3°; $F$ = 320 MV/m) at 1 ps; (b) and (c) show the 3-D distributions of space-charge density and exchange-correlation potential near the tip surface, respectively; (d) illustrates the calculated electron emission barrier profiles with and without including space charge fields (space charge potential and exchange-correlation effects) in JWKB approximation in front of tip surface by Eqs. (20) and (21), respectively; (e) the influence of a non-planar electron emitter on the calculated local electrostatic potential and the total electron emission energy barrier profile.

Since the space charge fields can significantly affect the electron emission process locally at the apex of Cu nano-tip, it would be interesting to further track the structural evolution of nano-emitter under such circumstances. By performing the ED-MD simulations with and without space charge fields, the atomic structure evolutions of Cu nano-tips under a static E-field (320 MV/m) are illustrated in Figures. 22(a) and 22(b), respectively. In Figure 22(c), the variation of the heigh of nano-tip is displayed for two different ED-MD simulations. The inclusion of space charge fields causes a relatively quick heating of Cu nano-tip. Within less than 90 ps, we observe a strong elongation and thinning of nano-tip. Between 90 ps and 120 ps, the thermal runaway event occurs, and the Cu nano-tip is blunted (See Figure 22(a)). On the other hand, without considering space charge fields, we only see a strong deformation of Cu nano-tip, but the temperature could not rise above 2000 K. The Cu nano-tip is also blunted because the thermal conductivity of Cu is high, leading to a rapid crystallization process in the molten region. However, thermal evaporation event is not seen in the whole ED-MD simulation which lasts about 150 ps. To further verify the mechanism that the enhanced electron emission process drives a rather rapid temperature raising phenomenon due to the Joule heating process in Cu nano-tip after including the space charge fields, the variations of total emission current versus time are plotted in Figure 22(b). Indeed, we find ED-MD simulation with space charge fields predicts higher total emission current than that of emission process without those effects for Cu nano-tip, excluding the profiles for blunted geometries between 110 ps and 150 ps. Overall, we demonstrate the importance of incorporating the space charge fields especially the exchange-correlation potential in calculating the field emission current for micro-protrusion or nano-emitter in the future ED-MD simulations.

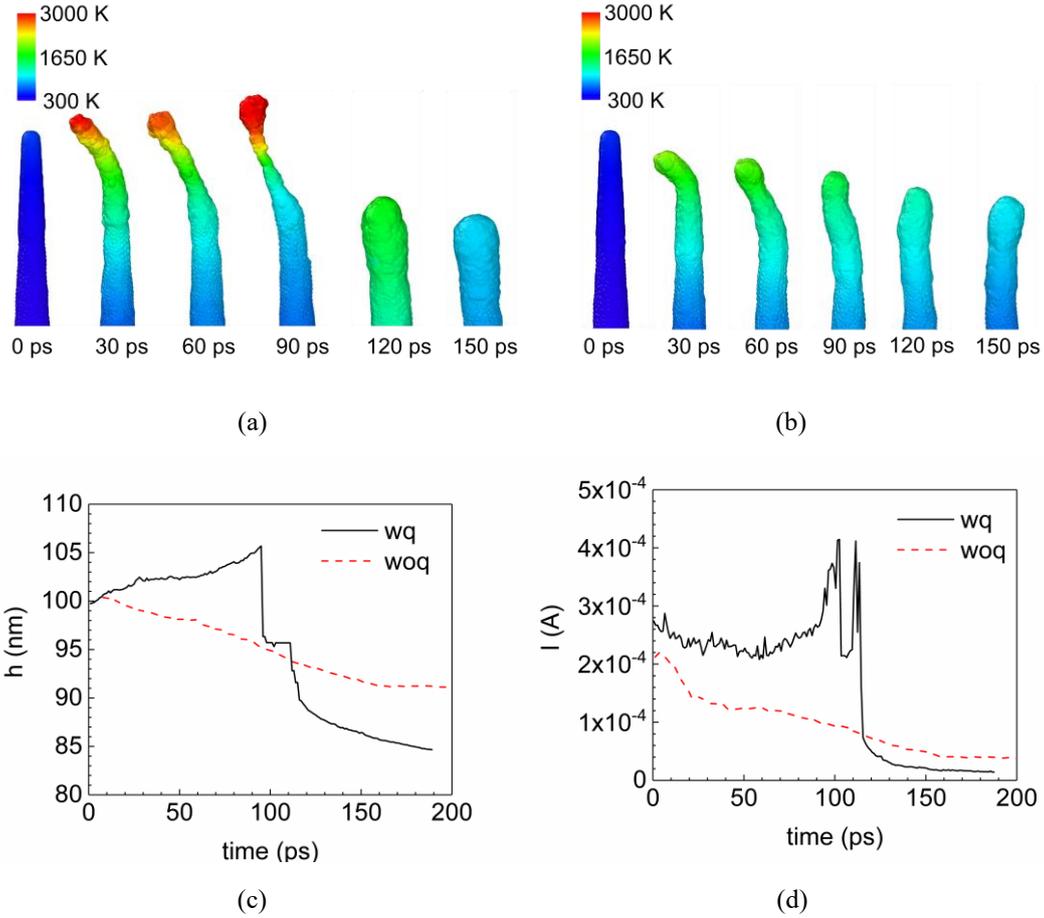

**Figure 22** The atomic structural evolutions of Cu nano-tip ($h_0 = 100$ nm, $r_0 = 1$ nm, $\theta_0 = 3°$; $F = 320$ MV/m) with and without considering space charge fields (Space charge potential and exchange-correlation effects) in (a) and (b), respectively; (c) The variations of the height of Cu nanotip and the total emission current versus time profiles of Cu nano-tip (d).

### 5.3 Nano-protrusions on the surface of planar W-Mo metal electrode

The formation of micro-protrusions on the planar metal electrode could be initiated from the nano-protrusions. The initiation and growth of nano-protrusions on the Cu polycrystalline metal surface have been attributed to the surface atomic diffusion mechanism along the grain boundary, triggered by the electric field stress or Maxwell stress [54]. During the formation of nano-protrusions on metal surface, the field emission process may be very weak, but the surface morphology of planar electrode changes significantly. Our previous studies implied that the thermal evaporation of micro- or nano-protrusions caused by the intense electron emission process could lead to the electric pre-breakdown condition [27-30]. In this section, we apply FEcMD program to demonstrate the formation of nano-protrusions on a planar metal electrode made of equal-molar W-Mo alloy.

Before performing the molecular dynamics simulations in FEcMD program under E-field, we first prepare a polycrystalline equal-molar W-Mo alloys. The ATOMSK [72] was adopted to build the initial atomic structure for the polycrystalline W-Mo alloy using the Voronoi tessellation, and OVITO software [47] was used for both analyzing and visualizing the atomic structures. As can be seen from Figure 23(a), the initial atomic structure of equal-molar W-Mo alloy consist of 9 randomly oriented grains with dimensions 16 × 16 × 8 nm, and which contains 160000 atoms. For the initial atomic structure of polycrystalline W-Mo alloy, atoms with a distance less than 2.55 Å from other neighboring atoms must be removed from the atomic model, and which are the artifacts of Voronoi tessellation. Otherwise, this atomic structure refining step is important to stabilize the following molecular dynamics simulations for polycrystalline W-Mo alloy. Then, we employ FEcMD program to perform the standard structural relaxation for totally 200 ps under NPT ensemble by setting the temperature to 900 K, and time step size to 2 fs, allowing the mechanical stress to be fully relaxed, and atomic positions are thermally equilibrated. Finally, the fully relaxed polycrystalline equal-molar W-Mo alloy is transferred to a large rectangular box with the normal periodic boundary conditions in the lateral directions, but which is aperiodic in z-direction. The upper surface of W-Mo alloy is exposed to a long vacuum layer. Meanwhile, all atoms on the lower surface are fixed to their relaxed positions in the previous step. Eventually, the stable polycrystalline W-Mo surface is obtained. To identify surface atoms and nano-protrusion, the atomic coordination number analysis (CNA) is performed. The cut-off radius for CNA is set to 5.0 Å, and the threshold value of 37 is employed. More specifically, atoms with CAN < 37 are assigned as the surface atoms or nano-protrusion. Finally, the EAM potentials employed for W-Mo alloys are provided in Ref [73].

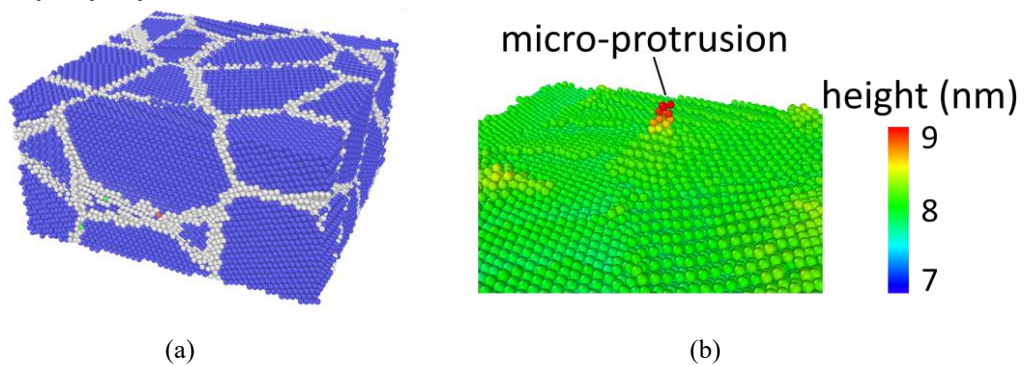

(a) (b)

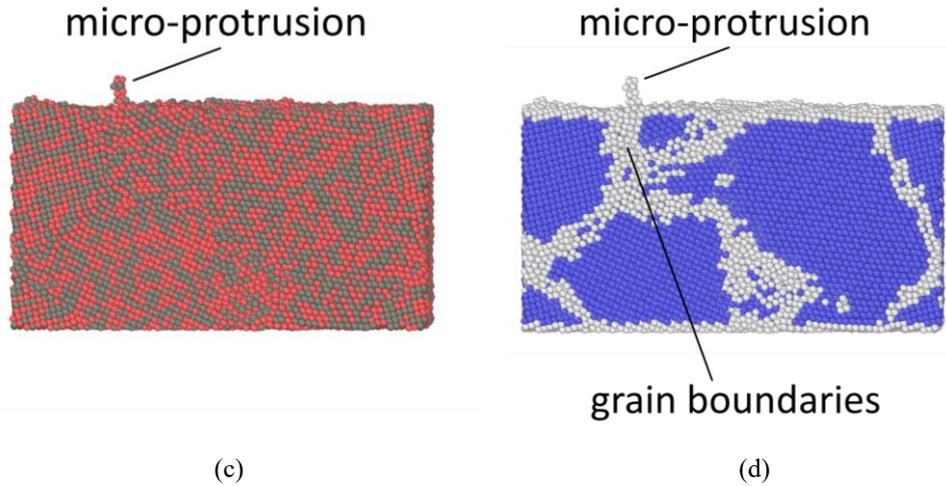

|  (c)  |  (d)  |

**Figure 23** The structural evolution of a 16 × 16 × 8 nm equal-molar polycrystalline W-Mo alloy structure with 9 randomly oriented grains under Maxwell stress rising at a rate of 0.024 GPa/ps for 880 ps: (a): the initial atomic structure. The blue and write atoms are BCC and grain boundary respectively, which is colored by polyhedral templating matching (PTM). The height map and side view of the micro-protrusion at 880 ps are shown in (b), (c) and (d). In (c), the red atoms are W, and the black atoms are Mo.

For the MD simulation under the E-field, the Maxwell stress is gradually increased with a constant rate of 0.024 GPa/ps during a duration of 880 ps. In Figures. 23(b) and 23(c), top and side views of a nano-protrusion are illustrated. In Figure 23(b), it is found that the height of the formed nano-protrusion is roughly equal to 9 nm. Furthermore, the chemical composition of nano-protrusion is also found to the W-Mo alloy. The segregation of W or Mo elements on the surface is not seen in the equal-molar W-Mo alloy under Maxwell stress in our MD simulation. Finally, the initiation and growth mechanism of nano-protrusion on planar equal-molar W-Mo alloy is implied in Figure 23(d), consisting with previous study that the grain boundary plays the vital role in the formation of nano-protrusion on planar metal electrode.

**5.4 Structural evolutions of Cu and W-Mo nano-tips with machine learning potentials**

The machine learning potential (MLP) provides the unique advantage that it could be conveniently generated using highly efficient muti-variable linear regression algorithm implemented in many open-source libraries by simply employing the atomic structure, energies, forces, and stresses obtained from standard first-principles molecular dynamics simulations as the training and validation data sets. Here, we demonstrate the use of MLP in FEcMD program to study atomic structure evolution of nano-tips made from Cu and equal molar W-Mo alloys under high E-field.

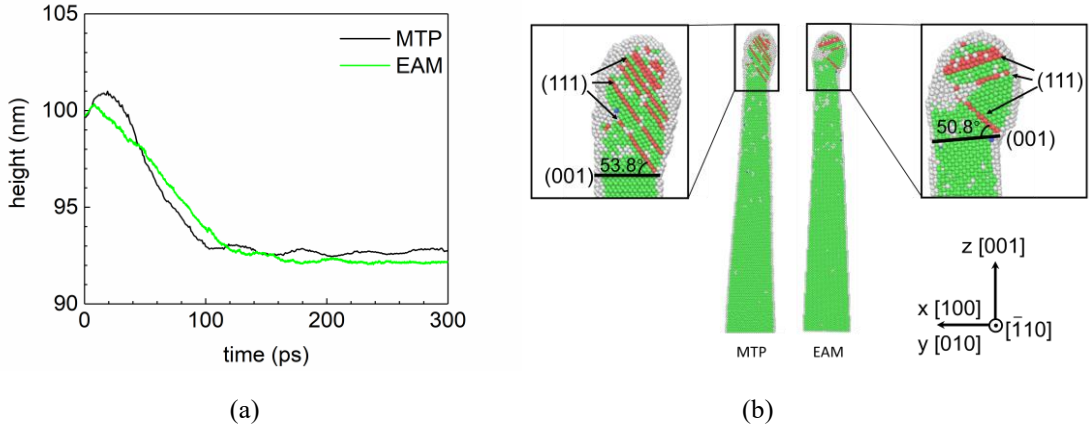

(a)  (b)

**Figure 24** ED-MD-PIC simulations of Cu nano-tip ($h_0$ = 100 nm, $h_0$ = 1nm and $\theta_0$ = 3°) using both EAM and MTP methods under static E-field of 300 MV/m: (a): the variations of height versus time; (b): atomic structure of Cu nano-tip analyzed by PTM (green: FCC, red: dislocation line) at 200 ps.

The machine learning potentials employed in this work are referred to the moment tensor potentials (MTPs) [43]. To generate the MTPs for Cu and W-Mo alloys, the open source MLIP library is used [43]. Otherwise, the training and validation data sets for MTPs are obtained from the first-principles molecular dynamics simulations using VASP program [74]. The training data sets include the atomic configurations, energies, atomic forces, and stresses provided by conducting the FPMD simulations at various temperatures and with different lattice parameters. The bootstrapping technique is adopted for obtaining MTPs through the active learning algorithm [43]. More specific details regarding the preparation of training data sets using MLIP library and VASP program are provided in Appendix E.

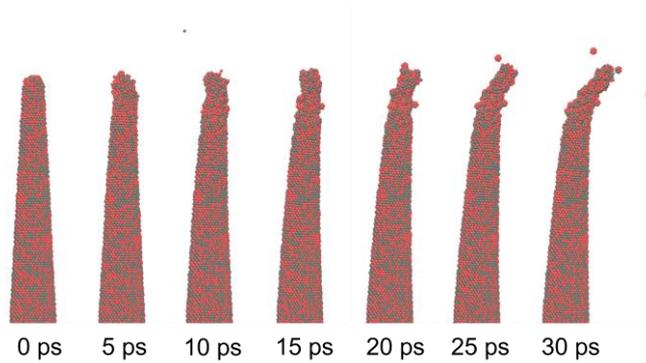

**Figure 25** Atomic structure evolution of the equal-molar W-Mo alloy emitter ($h_0$ = 100 nm, $r_0$ = 1nm and $\theta_0$ = 3°) under a constant electric field (F = 500 MV/m) using MTPs. The red spheres are W, and the black ones are Mo.

First, we conduct the ED-MD-PIC simulations for conical Cu nano-emitter ($h_0$ = 100 nm, $r_0$ = 1nm and $\theta_0$ = 3°) under a static E-field and MTP. The variation of the tip height during the ED-MD simulations is displayed in Figure 24(a). The height versus time profile of MTP is also compared with that of EAM potential. Overall, we see good agreement between the two interatomic potentials for the obtained height profiles. Nevertheless, the elongation of Cu nano-tip at the beginning of ED-MD-PIC simulation using MTP is found to be significant, and which also proceeds slowly, compared to that of EAM potential. Otherwise, both MTP and EAM potentials predict that the height of Cu nano-tip is eventually stabilized after 100 ps. The structural deformation mechanism of Cu nano-tip is further illustrated in Figure 24(b) using both MTP and EAM potentials. The titling of nano-emitter is mainly caused by the propagation of dislocations in the interior of the structure [28]. Due to the high resistively heating at the apex of Cu nano-emitter induced by intense electron emission process, the tip of nano-emitter is partly in a molten state, as shown in Figure 24(b) by all white atoms.

Next, similar ED-MD-PIC simulations are carried out for equal-molar W-Mo alloy using MTPs. In this case, the static E-field strength is increased to 500 MV/m. From the previous studies, it is anticipated that the thermal evaporation event could occur under such high E-field for pure W or Mo nano-tips of similar size [27]. In Figure 25, the structural evolution of equal-molar W-Mo nano-tip is shown. Obviously, we observe the evaporation of atomic clusters on the tip of W-Mo nano-emitter after running ED-MD-PIC simulation for 25 ps. Multiple small atomic clusters are detached from W-Mo nano-emitter at about 30 ps. For a better understanding of the atomic structure evolution of W-Mo nanotip during the ED-MD-PIC simulation, the radial distribution functions (RDFs) are calculated at the apex region, and the results are shown in Figures 26(a)-(c) for all atomic pairs (W-W, W-Mo, and Mo-Mo pairs), W-W pair and Mo-Mo pair, respectively. Initially, we observe multiple shark spikes in the obtained RDF profiles for different atomic pairs, which clearly indicate the ordering of atomic structures at the beginning of simulation. As the ED-MD-PIC simulation proceeds, all RDFs are significantly smeared, and many sharp spikes at the large distance are completely disappeared, indicating the apex region of W-Mo nanotip goes from the ordered state to the disordered configuration. The disorder in the atomic structure is obviously related to the resistive heating process induced by strong electron field emission mechanism. Strong resistive heating can directly melt the apex region of W-Mo nanotip, and segregation of W or Mo atoms may also occur. Here, we calculate the atomic short-range order (SRO) parameter for W-Mo nanotip as a function of MD duration, as shown in Figure 26(d). The Warren-Cowley SRO form is adopted, and which is given by Eq. (31).

$$\alpha_{\mathrm{W}} = 1 - \frac{z_{\mathrm{Mo}}}{z_{\mathrm{tot}}(1-c_{\mathrm{W}})} \tag{31}$$

In Eq. (31), $\alpha_W$ represents the SRO within an atomic shell centered at W atom, $Z_{Mo}$ and $Z_{tot}$ are the number of Mo atoms and the total number of atoms (W+Mo) surrounding a W atom within the same shell, and $c_W$ denotes the concentration of W in the W-Mo binary alloy. We use a cutoff radius of 3.815 Å, which is the average of the next-nearest and second-nearest atomic distances, for the W atomic shell in the calculation. According to Eq. (31), for a completely segregated W-Mo atomic structure, SRO=1, while for a completely disordered solid solution crystal structure, SRO=0. A higher value of SRO indicates a more significant aggregation of W atoms in the W-Mo crystal structure. In Figure 26(d), the SRO of the W-Mo alloy nano-tip decreases slowly from an initial value of 0.285 to 0.26, indicating that during the melting process induced by field emission, W atoms prefers to form a disordered solid solution with Mo atoms in W-Mo alloy. Surface segregation of one of the constituting elements is not seen for equal-molar W-Mo alloy. Interestingly, the formation of nano-protrusions on a planar electrode of equal-molar W-Mo alloy is addressed by performing MD simulations using EAM potentials in section 4.3, segregation is again not seen in that case.

The chemical composition of the evaporated atomic clusters can be directly inspected from Figure 25. All escaping atomic clusters are made of W and Mo atoms, indicating that the chemical bonding strength between W and Mo is comparable to that of either W-W or Mo-Mo. This conclusion is further confirmed by calculating the atomic percentage of W atoms in the evaporates, as shown in Figure 26(e). Clearly, the evaporates also contain the equimolar Mo and W atoms, the same as that of W-Mo nanotip. Therefore, evaporation of equimolar W-Mo alloy is congruent under the intense resistive heating due to the electron field emission.

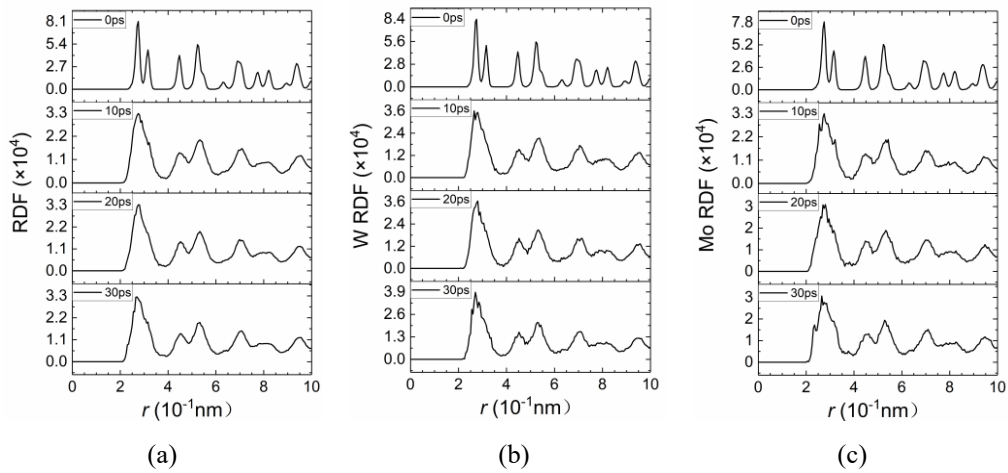

(a)          (b)          (c)

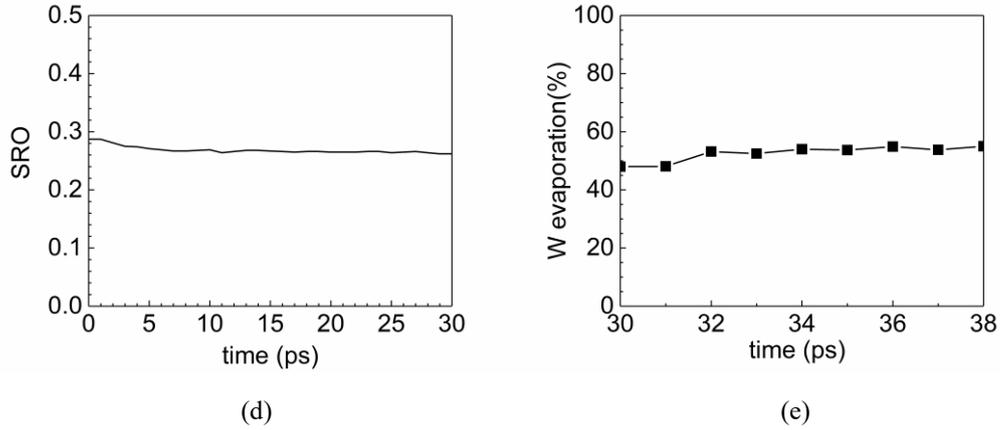

(d)        (e)

**Figure 26** Atomic structure analysis at the apex for metal nanotip ($h_0$ = 100 nm, $r_0$ = 1nm and $\theta_0$ = 3°) consisting of the equimolar W-Mo alloy under F = 500 MV/m using FEcMD software: (a) Radial distribution function (RDF); (b) RDF for W-W pair; (c) RDF for Mo-Mo pair; (d) Atomic short-range order (SRO) parameter; (e) Atomic percentage of W atoms in the evaporates.

**5.5 E-field simulations for different nanotip geometries**

The ED module in FEcMD program can be used independently to perform the advanced multi-physics simulations for micro- and nano-protrusions or electron emitters based on finite element method. This feature is very useful for a quick evaluation of the electron emission characteristics of metal electron emitter when the physical properties and geometry are specified. FEcMD program relies on a self-built library to generate various shapes for electron emitter, including the hemisphere, cylinder, pyramidal-hemisphere [20], prolate-spheroidal [75], hemi-ellipsoidal [76] and mushroom-head [27]. Each geometry is specified by a set of parameters. FEcMD program provides a script for users to generate the protrusions with different geometries and sizes. Then the geometry is imported into the FEMOCS library to produce the finite element mesh for multi-physics simulation. In the Figure A1 and Table. A1 of Appendix A, the supported geometries and their featured parameters are summarized. Additionally, an example of the script for generating the tip geometry is also provided in Appendix A.

Here, we conduct multi-physics simulations for three different electron emitter geometries using FEcMD program, and the results are illustrated in Figure 27. In Figures. 27(a)~(c), the electric field vectors and temperature distribution are shown for prolate-spheroidal, hemi-ellipsoidal and mushroom-head emission tips at different simulation times, respectively. Those three geometries exhibit a relatively sharp apex region, resulting in the high local electric field strength and strong electron emission process.

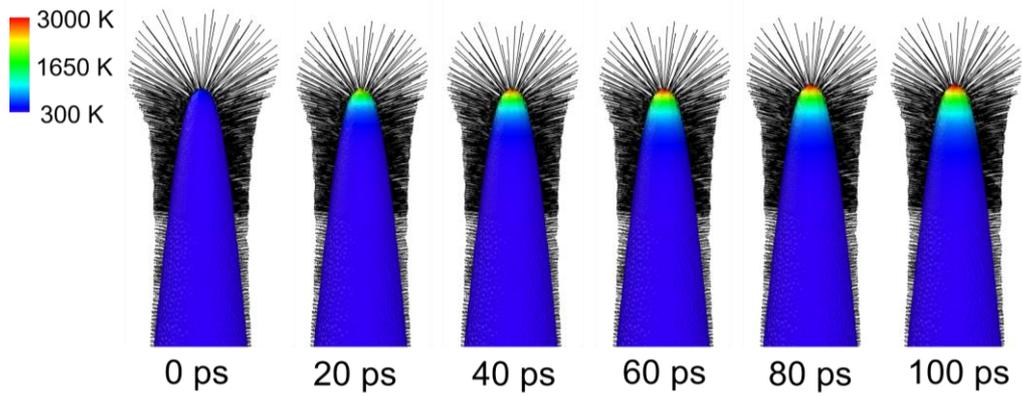

(a)

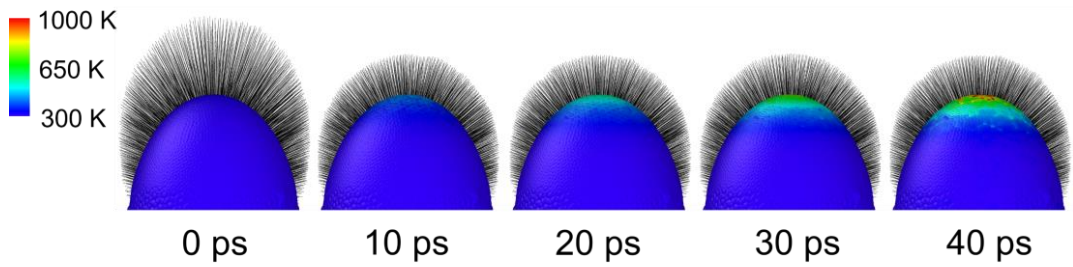

(b)

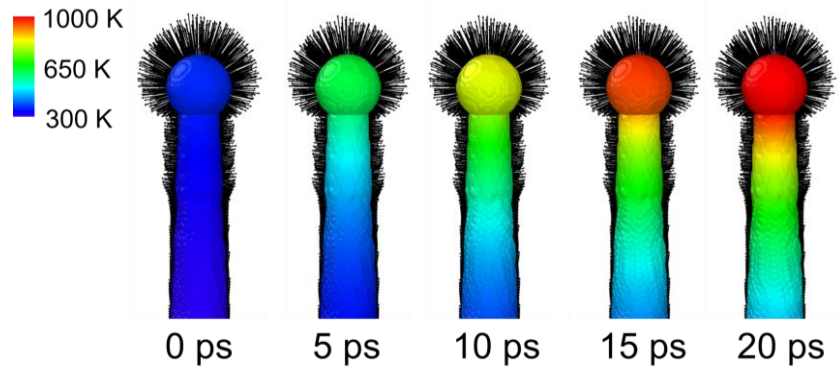

(c)

**Figure 27** The multi-physics simulations of electron emitters with various geometries for electric field and temperature distributions: (a): the prolate-spheroidal emission tip ($r$ = 5 nm, $h$ = 1 μm, $\theta$ = 4°) under a constant electric field (600 MV/m); (b): the hemi-ellipsoidal emission tip ($r_b$ = 1.5 μm, $h$ = 2.5 μm) under a constant electric field (4 GV/m). (c): the mushroom-head emission tip ($r_t$ = 1.5 nm, $R$ = 3 nm, $h$ = 100 nm, $\theta$ = 3°) under a constant electric field (400 MV/m).

Without invoking the MD module, the FEcMD software can be employed to study the field emission properties and even the electrical pre-breakdown E-field for micro-emitter or micro-protrusions. For micro-size electron emitters in a conventional field emission device, the distance between anode and cathode is at least several hundred micrometers. Under such conditions, the space charges can be treated as classic particles and the space charge effects are negligible in solving the WKBJ model to obtain the electron tunneling barrier height. By performing the standard finite-element simulation within ED and HC modules in FEcMD software, the pre-breakdown E-field or the time to thermal runaway can be estimated from the calculated phonon temperature of electron emitter. For a demonstration, the ED+HD simulations are conducted for the electron emitter made of Cu pyramidal-hemisphere with total height $h = 1$ μm, radius of hemispherical cap $r = 10$ nm, and half-angle $\theta = 5°$. The applied E-field is set to 300 MV/m, and the anode-to-cathode distance of 100 μm. The anode has a planar geometry. All simulations are carried out without including space charge effects (exchange-correlation potential and space charge potential). Notably, the atomic structure of this size for Cu electron emitter contains more than $10^7$ atoms, and the task within ED-MD-PIC scheme could take many weeks to complete using multiple parallel threads on two CPU processors. In Figure 28, the obtained phonon temperature, E-field and emission current density distributions and their evolutions with simulation time are illustrated. Figure 8(a) shows the concentration of local E-field at the apex of Cu electron emitter. As a result, the obtained electron emission current density gives large values at the same region, as shown in Figure 8(b). The calculated local field enhancement factor at the apex of the emitter is found to be 59.3. As can be seen from Figure 8(c), the two-temperature model reveals that the phonon (lattice) temperature is significantly smaller than that of electron temperature at the beginning of simulation, indicating the heat transfer mechanism dominated by hot electrons. However, the presence of electron-phonon coupling allows the phonon to dissipate heat efficiently. Eventually, the phonon and electron temperatures approach each other after 40 ps. Interestingly, the finite element simulations also show that both the temperature and total field emission current (See Figure 28(d)) are stabilized to some constants using ED+HC modules in FEcMD software. Obviously, the results are not realistic for the simulated electron emitter because the phonon temperature at the apex is well above the melting point of Cu shortly after the running the simulation for 20 ps (See Figure 8(a)). Within the full ED-MD-PIC simulation, such a high phonon temperature already causes the melting of the tip region, and thermal runaway could occur due to the strong structural deformation under the high electric stress. Therefore, the pyramidal-hemispherical Cu electron emitter of this size ($h = 1$ μm, $r = 10$ nm, and $\theta = 5°$.) may subject to the pre-breakdown condition in the atomistic simulation using ED-MD-PIC method.

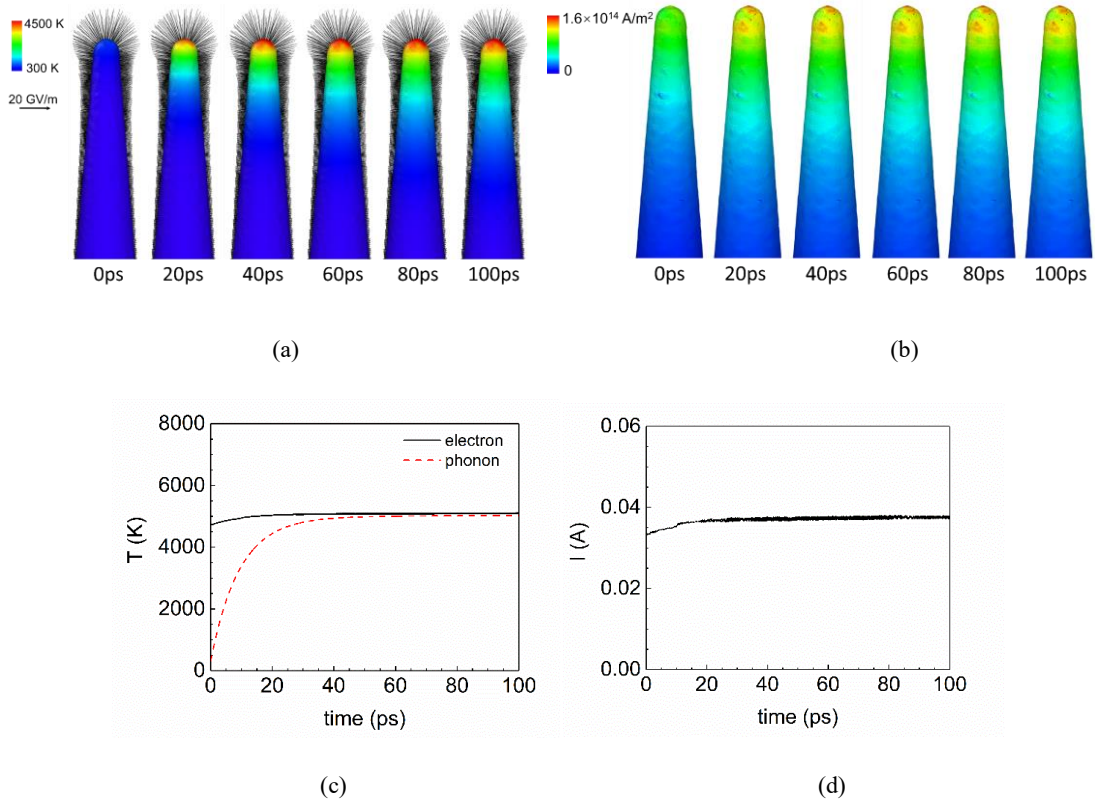

Figure 28 The finite element simulations of E-field and thermal conduction for the pyramidal-hemispherical Cu electron emitter using ED+HC module in FEcMD software under the applied F = 300 MV/m: (a) Temperature and E-field distributions; (b) Field emission current density distribution; (c) Phonon and electron temperature profiles; (d) Total electron emission current profile. The dimensions of electron emitters are $h = 1$ μm, $r = 10$ nm, and $\theta = 5°$.

## 6. Conclusions

In this paper, we presented a multi-physics and multi-scale simulation methodology for investigating and understanding electron characteristics and atomic structure evolutions of micro- and nano-protrusions in vacuum using FEcMD software package. Although, our new implementations of ED-MD simulations in FEcMD program were based on the already existing open-source libraries and methods, it indeed provided some vital updates in the algorithms to support freshly new physical insights for the field emission process dynamically coupled with atomic structure changes, including the two-temperature heat conduction mechanism, space charge fields (space charge potential and exchange-correlation effects), and the interface to EAM and MTP methods for multi-component alloys. We believe those major upgrades available in FEcMD software package could not only provide a powerful computational methodology to study the vacuum electric pre-breakdown mechanism induced by the hypothesized thermal evaporation of micro- or nano-protrusions on the metal electrodes, but also greatly benefits the design of modern nanoelectronics and nano-emitters.

**Acknowledgements**

This research is financially supported by the Young Talent Support Plan at Xi'an Jiaotong University awarded to Bing Xiao with the contract No: DQ1J009. The authors would like to thank Prof. Flyura Djurabekova (Helsinki Institute of Physics and Department of Physics, University of Helsinki), Prof. Andreas Kyritsakis (Institute of Technology, University of Tartu), Dr. Mihkel Veske (Helsinki Institute of Physics and Department of Physics, University of Helsinki) for providing us the FEMOCS code. Bing Xiao also would like to thank Dr. Sergio Catatroni (CERN, European Organization for Nuclear Research) for his valuable comments on this work.

**Appendix A: Generate different geometries of nanotips**

The FEcMD software provides the scripts for the users to build the nanotips with six different geometries, including Pyramidal-hemisphere, Mushroom-head tip, Prolate-spheroidal, Hemi-ellipsoidal, Cylinder and Cone. In Figure A1 and Table A1, the schematic diagram of all geometries along with the corresponding analytical expressions are provided.

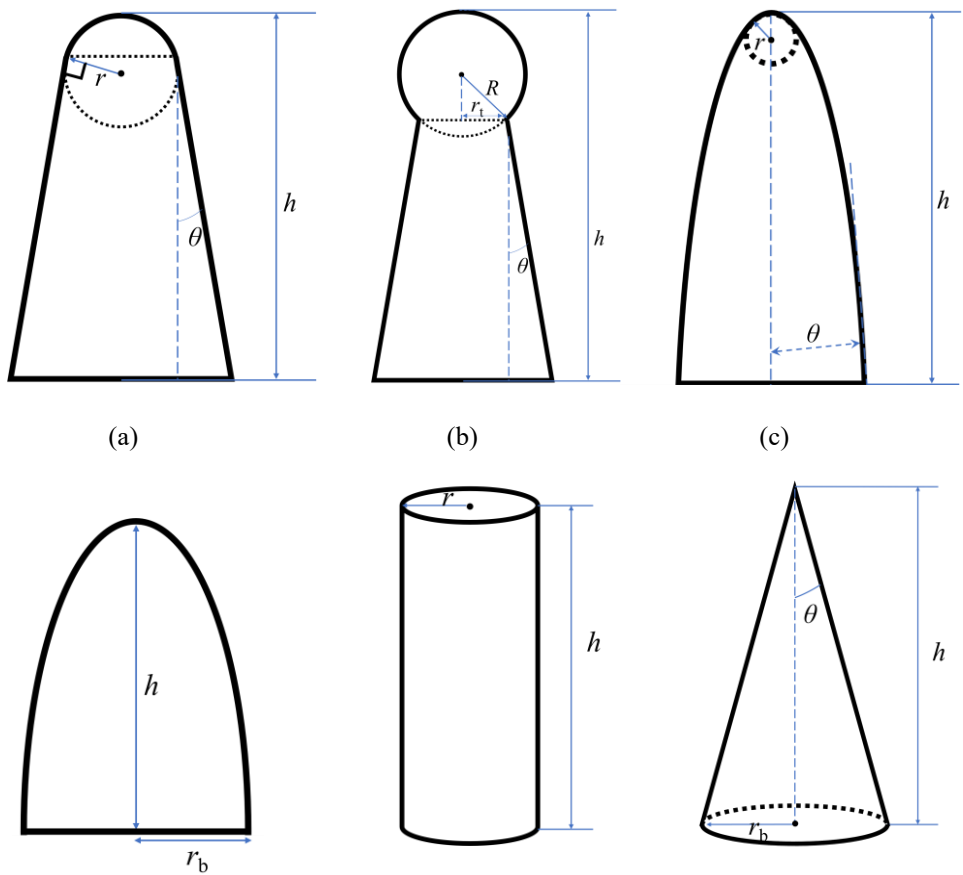

(a)　　　　　　　(b)　　　　　　　(c)

(d) (e) (f)

**Figure A1** Schematic diagrams of various types of emitter tips that can be built by the MD module and parameters determined by users: (a): the pyramidal-hemisphere emission tip ; (b): the mushroom-head emission tip ; (c): the prolate-spheroidal emission tip; (d): the hemi-ellipsoidal emission tip; (e): the cylinder emission tip; (f): the cone emission tip.

**Table A1** Types of emitter tips that can be built by the MD module, corresponding expressions and parameters determined by users.

| Emitter tips | Expressions | Parameters |
|---|---|---|
| (a) Pyramidal-hemisphere | $\begin{cases} x^2 + y^2 + (z-h+r)^2 = r^2, & z \geq h - r(1-\sin\theta) \\ \sqrt{x^2+y^2} + z\tan\theta = \\ r\cos\theta + \tan\theta[h - r(1-\sin\theta)], & 0 \leq z < h - r(1-\sin\theta) \end{cases}$ | Total height $h$, Radius of the hemispherical cap $r$, Half-angle $\theta$ |
| (b) Mushroom-head tip | $\begin{cases} x^2 + y^2 + (z-h+R)^2 = R^2, & z \geq h - R - \sqrt{R^2 - r_t^2} \\ \sqrt{x^2+y^2} + z\tan\theta = \\ r_t + \tan\theta\left(h - R - \sqrt{R^2 - r_t^2}\right), & 0 \leq z < h - R - \sqrt{R^2 - r_t^2} \end{cases}$ | Total height $h$, Radius of the mushroom cap $R$, Radius of cone top $r_t$, Half-angle $\theta$ |
| (c) Prolate-spheroidal | $\begin{cases} x = a\sinh(u)\sin(v)\cos(\varphi) \\ y = a\sinh(u)\sin(v)\sin(\varphi), \\ z = a\cosh(u)\cos(v) \end{cases} \begin{pmatrix} 0 \leq u < u_{max} \\ 0 \leq \varphi < 2\pi \end{pmatrix}$ <br><br> the half of the foci distance $a = \dfrac{r}{\sin\theta \tan\theta}$, <br><br> $v = \pi - \theta$, <br><br> $t = \dfrac{a\cos(v) - h}{a\cos(v)}$, $u_{max} = \log\left(t + \sqrt{t^2 - 1}\right)$, <br><br> The distance from the tip top to anode $d = a\cos\theta$ | Total height $h$, Tip radius $r$, Half-angle $\theta$ |
| (d) | $\begin{cases} x = r_b \sin(\theta)\sin(\varphi) \\ y = r_b \cos(\theta)\sin(\varphi), \\ z = h\cos(\varphi) \end{cases} \begin{pmatrix} 0 \leq \theta < 2\pi \\ 0 \leq \varphi < \dfrac{\pi}{2} \end{pmatrix}$ | Total height $h$, |

| | | | |
|---|---|---|---|
| Hemi-ellipsoidal | | | Radius of the bottom $r_b$ |
| (e) Cylinder | $x^2 + y^2 = r$, $0 \leq z \leq h$ | | Total height $h$, Radius of the bottom and the top $r$ |
| (f) Cone | $\sqrt{x^2 + y^2} + z \tan\theta = r_b$, $0 \leq z \leq h$ $r_b = h\tan(\theta)$ | | Total height $h$, Radius of the bottom $r_b$ |

To obtain the atomic coordinates file "mdlat.in.xyz" for each geometry of emission tip, user needs to provide two input files separately in a target folder:

(1) Unit cell of the bulk metal in the file "CONTCAR" for building a specified geometry tip;

(2) Tip parameters in the input file "md.in".

The CONTCAR file defines the lattice structure of bulk metal in terms of lattice vectors, atomic coordinates, types of elements and number of atoms, and its format is precisely the same as that of the file employe in standard VASP code.

**Appendix B.** Input and output files for molecular dynamics.

To run the stand-along molecular dynamics simulation, the MD and AF modules need to be executed simultaneously. The following three input files must be prepared before the execution:

(1) The atomic coordinates file "mdlat.in.xyz" for the modelling structure.

(2) The parameters for L-J potentials or potential files for EAM and MTP methods:

   (a) LJ potential parameters in "md.in" for each atomic pairs;

   (b) The EAM potential files in the folder "potential/";

   (c) The machine learning potential files in the folder "potential/";

(3) The key tags for MD simulation in file "md.in":

```
1    deltaT = 4              # The time interval of each step [fs]
2    temperature = 300       # Initial temperature [K]
```

| 3 | Nelems = 1 | # The number of elements |
|---|---|---|
| 4 | elems = Cu | # Elements type |
| 5 | mass = 63.546 | # Relative atom mass for each element |
| 6 | stepAvg = 10 | # The average steps of statistical energy and temperature |
| 7 | stepMovie = 100 | # The average steps of the output atomic trajectory |
| 8 | stepEquil = 1000 | # Total time steps for initial optimization |
| 9 | stepLimit = 100000 | # Total number of simulation steps |
| 10 | randSeed = 7 | # Random seed |
| 11 | boundary = p | # Boundary condition: periodic (p) or nonperiodic (n) |
| 12 | structure_type = FCC | # Structure type: FCC, BCC or cubic |
| | | # (consistent with the input atomic structure file) |
| 13 | lattice = 3.6147 | # Lattice parameters [Angstrom] |
| 14 | interact_method = nebr | # Statistical method of interacting atom pairs: |
| | | # allpairs, cell or nebr |
| 15 | force_type = MTP | # Types of atomic interactions: |
| | | # lj, metal, alloy, snap, eamfs, MTP |
| 16 | force_file = in/potential/Cu.mtp | # Atomic interaction potential file name |
| 17 | ensemble = NVT | # Ensemble: NVE, NVT, NPT or nonEquil |
| 18 | Nthreads = 8 | # Parallel threads for MD |
| 19 | pedestal_thick = 0 | # Fixed atomic thickness at the bottom |
| 20 | #tip Maker | |
| 21 | #-------------------------------------- | |
| 22 | cell_order = disorder | # Atomic ordering of the alloys constructed: |
| | | # order or disorder |
| 23 | cell_file = in/mdlat.in.xyz | # Field emitter atomic strcture file name |
| 24 | initUcell = 10 10 10 | # The size of cell expansion |
| 25 | tip_file = in/mdlat.in.xyz | # The generated field emitter atomic strcture file name |
| 26 | tip_R = 30 | # The radius of the mushroom cap R [Angstrom] |
| 27 | tip_r = 15 | # The radius of conical top rt [Angstrom] |
| 28 | tip_h = 500 | # Total height [Angstrom] |

```
29    tip_theta = 3              # Half-angle [degree]
```

During the execution of the molecular dynamics simulations, the main outputs are created into the folder "out/":

  md.movie: the trajectory file of the tip atoms during the evolution

  edhc.in.xyz： current atomic coordinates which output to ED

  summary.dat: atomic average energy, average temperature, tip height. etc

  T_profile.dat: the temperature profile obtained from MD simulation

  V_profile.dat: the atomic velocity profile

  rdf.dat: the radial distribution function

**Appendix C.** Input and output files for ED+HC simulations.

Running the ED and HC modules to calculate the electric field, field emission, and heat conductions in three-dimensional rigid nano-tips on finite element method, users need to prepare five input files. They are the finite element mesh points coordinate written in file "edhc.in.xyz" for field emission emitter, the user modified computational parameters in file "edhc.in", the built-in file "GetelecPar.in" for field emission calculation, the electrical resistivity given in file "rho_table.dat", and the phonon thermal conductivity file "phonon_kappa_table.dat":

(1) The finite element mesh file "edhc.in.xyz";

(2) The computational parameters file "edhc.in":

```
1    # General

2    infile = in/edhc.in.xyz     # mesh coordinates of emitters

3    timestep = 4                # time step (should equal to ΔtMD in ED-MD simulation) [fs]

4    timelimit = 200             # total simulation time [ps]

5    movie_timestep = 1000       # time interval to write the movie files; 0 turns writing off [fs]

6    radius = 70                 # inner radius of coarsening cylinder [A]

7    coarse_theta = 10           # apex angle of coarsening cylinder [deg]

8    coarse_factor = 0.5 12 2    # coarsening factor, larger number gives coarser surface
                                 # 1st - coarsening factor for atoms outside the coarsening cylinder
                                 # 2nd - minimum distance between atoms in nanotip below apex [latconst/4]
```

line 3: timestep = 4 # time step (should equal to $\Delta t_{MD}$ in ED-MD simulation) [fs]

|   |   |   | # 3rd - minimum distance between atoms in nanotip apex [latconst/4] |
|---|---|---|---|

10　# Cluster extraction processing

11　nnn = 12　　　　　　　　　# nr of nearest neighbours of bulk material within coord_cutoff radius; needs to be adjusted if coord_cutoff is changed

12　coord_cutoff = 3.1　　　　　# atomic coordination number analysis cut-off radius [A]

13　cluster_cutoff = 4.2　　　　# cluster anal. cut-off radius [A]; if 0, cluster anal. uses coord_cutoff instead

14　# Parameters for extending simulation domain

15　extended_atoms = in/extension.xyz　# file with atoms of extended surface

16　box_width = 10　　　　　　# minimal simulation box width [tip height]

17　box_height = 10　　　　　　# simulation box height [tip height]

18　bulk_height = 20　　　　　　# bulk substrate height [latconst]

19　# Field emission parameters

20　work_function = 4.59　　　　# work function [eV]

21　emitter_qe = true　　　　　#if true quantum effect of space charge will be considered

22　func_X_id = 1　　　　　　　#exchange functional id (Refers to Libxc.library)

23　func_C_id = 9　　　　　　　#correlation functional id (Refers to Libxc.library)

24　# Heating parameters

25　temperature_mode = double　#single (lattice heat conduction) or double (two-temperature model)

26　t_ambient =   300.0　　　　# temperature of bulk reservoir of simulation cell [K]

27　lorentz = 2.e-8　　　　　　# Lorentz constant [W Ohm K-2]

28　heat_cp = 3.491e-24　　　　# Volumetric heat capacity [J/(K*Ang^3)]

29　Ncell = 3 3 100　　　　　　# the size of temperature rescaling cells

30　heat_dt = 40　　　　　　　# heat calculation time step [fs]

```
31  # Field parameters
32  elfield_mode = DC           # Types of the applied electric field; DC, AC, pulse, or index E*(1-exp(-t/tau))
33  elfield = -0.0320000        # DC--value of applied electric field; pulse--amplitude value   [V/Angstrom][10000 MV/m]
34  anode_bc = neumann          # boundary condition type at anode; Dirichlet or Neumann
35  # PIC parameters
36  pic_dtmax = 0.5             # maximum time interval between particle collisions in PIC [fs]
37  electron_weight = 0.01      # electron superparticle weight [nr of SPs]
```

(3) the built-in file "GetelecPar.in" for field emission;

(4) The resistivity file "rho_table" format (example for the copper in nanoscale):

(5) The phonon thermal conductivity file "phonon_kappa_table.dat".

It is worth mentioning that if users want to enable or disable the coupled calculations of thermal diffusion, PIC, space charge, and quantum effects, you can set the relevant parameters in the "edhc.in" file.

The main outputs of ED-HC simulations are written in the following files:

➢ finite-element mesh generation results

  atomreader.ckx: the atomic coordinates read by FECMOS library

  surface_dense.xyz: the coordinates of the extracted tip surface points

  surface_coarse.xyz: the coordinates of the coarse-grained tip surface points

  mesh/*vtk: meshes generated for finite element calculation

➢ ED results

  ch_solver.movie: electric current and heat solver results during the evolution

  ch_solver.xyz: current and heat solver results currently being simulated

  E0.dat: the value of applied electric field during the evolution

  electrons.movie: the electron trajectory simulated by the PIC during the evolution

  electrons.xyz: the electron trajectory simulated by the PIC currently being simulated

   emission.dat: the electron filed emission data

   exchange-correlation.movie: exchange-correlation potential during the evolution

   exchange-correlation.xyz: exchange-correlation potential currently being simulated

   fields.movie: the electric field distribution on the tip during the evolution

   fields.xyz: the electric field distribution on the tip currently being simulated

   surface_fields.movie: the electric field distribution on the tip surface during the evolution

   surface_fields.xyz: the electric field distribution on the tip surface currently being simulated

   FJI.dat: the electric field, current density, and current data

   forces.movie: the force induced by electric field during the evolution

   forces.xyz: the force induced by electric field currently being simulated

➢ HC results:

   surface_temperature.movie: the electron temperature distribution on the tip surface during the evolution

   surface_temperature.xyz: the electron temperature distribution on the tip surface currently being simulated

   temperature_phonon.movie: the electron and phonon temperature distribution of the tip during the evolution

   temperature_phonon.xyz: the electron and phonon temperature distribution of the tip currently being simulated

   temperature.dat: the electron and phonon temperature data.

Among them, files with the extension ".movie" or ".xyz" can be visualized using software like OVITO software, while files with the ".vtk" extension can be visualized using Paraview software.

**Appendix D.** Input and output files for ED-MD-PIC simulation

For performing a standard ED-MD-PIC simulation, the required input and output files are a combination of those detailed in Appendix B and C. During the simulation, the atomic coordinates of the molecular dynamics simulation are automatically converted into the finite element mesh point file "edhc.in.xyz". Therefore, users have no need to provide this file separately. The parameter of "ensemble" in "md.in" should be set to "nonEquil" for non-equilibrium molecular dynamics of ED-MD-PIC simulations. Additionally, due to the consecutive movements of atoms in nanotip during MD simulation, it is computationally expensive to constantly update the finite element mesh at every step. Therefore, we have implemented a strategy where we track the root-mean-square distance (RMSD) that the atoms have moved since the last full finite-element calculation [21]. When it exceeds a predefined threshold, the finite element mesh is regenerated according to the new atomic structure of nanotip. Otherwise, if RMSD is below the threshold value, the existing mesh is used for electric field calculations. This approach significantly improves computational efficiency without losing the accuracy.

$$\text{RMSD} = \sqrt{\frac{1}{N}\sum_{i\in N}\left|\vec{r}_i - \vec{r}_i^{\text{ref}}\right|^2} \tag{D1}$$

Here, $N$ is the atoms number, and the accumulation term is the square of the distance between the current atomic coordinates $\vec{r}_i$ and the reference atomic coordinates $\vec{r}_i^{\text{ref}}$ that are used to construct the finite element mesh in the previous iteration. In that case, the user needs to set the threshold value for the RMSD in the file "fecocs.in". The value for the RMSD is determined empirically. It determines the interval at which the mesh rebuilding is triggered based on the accumulated displacement of the atoms.

    distance_tol = 0.45        # max RMS distance atoms are allowed to move between runs before the solution is recalculated; 0 enforces the mesh generation for every time-step

During the execution of the ED-MD-PIC model, dynamic outputs of all modules' results are stored in the "out" folder, refer to Appendix B and C for more details.

**Appendix E. Procedures for training of machine learning potentials**

The MD module with the open-source MILP library is employed to train a MTP for Cu. Similar to EAM potential, the use of MTP allows us to perform the ED-MD-PIC simulations for the metal nanotips containing up to $10^6$ atoms on two CPU processor. First, we calculate energies, atomic forces and stresses of various structures of Cu supercell model within 108 atoms by using VASP [74] as the training data set. Bootstrapping iteration technique is adopted for active learning of MTP, which is often called learning on-the-fly. We chose an MTP of level 16, where the cut-off radius is 5.5 Å, the minimum distance is 2 Å, and the radial base size is 8. It is important to note that the quality of the machine learning potential is not only related to the functional form, but also determined by the quality of the training set obtained in an appropriate query strategy. In order to efficiently obtain a high-quality training set, the active learning bootstrapping iterations [43] is used.

After training the machine learning potential, we recalculate lattice constants, elastic constants, single vacancy defect formation energy, and melting points as shown in Table E1 by running the MD module alone. All calculated results are consistent with the experimental values of copper.

**Table E1** Properties of copper predicted by the MTP potential in comparison with experimental data

| properties | experiment | MTP |
|---|---|---|
| $a_0$ (Å) | 3.615[a] | 3.633 |
| $c_{11}$ ($10^{11}$ Pa) | 170.0[b] | 170.9 |
| $c_{12}$ ($10^{11}$ Pa) | 122.5[b] | 119.6 |
| $c_{44}$ ($10^{11}$ Pa) | 75.8[b] | 76.6 |
| $E_v$ (eV) | 1.27[c] | 1.09 |
| $T_m$ (K) | 1368[d] | 1350 |

[a]Reference [77].
[b]Reference [78].
[c]Reference [79].
[d]Reference [80].

Following the similar steps and procedures of Cu case, we also train the MTP of W-Mo alloy by employing Bootstrapping iterations technique. We selected 39k crystal structures with up to 12 atoms of W-Mo supercell models with various W/Mo ratios, and different lattice types (FCC, BCC, HCP) reported from literature [81, 82] as the training data sets.

**Appendix F. Discretization of the Poisson equation**

The Poisson's equation (see Eq. (10)) is multiplied by the weighting function ω and integrated over the domain Ω to obtain:

$$\int_\Omega \omega \vec{\nabla} \cdot \left( \varepsilon \vec{\nabla} \varphi \right) d\Omega + \int_\Omega \omega \rho d\Omega = 0$$

By utilizing the properties of differentiation and the divergence theorem, we can derive:

$$\oint_\Gamma \omega \varepsilon \vec{\nabla} \varphi \cdot \vec{n} d\Gamma - \int_\Omega \varepsilon \vec{\nabla} \omega \cdot \vec{\nabla} \varphi d\Omega + \int_\Omega \omega \rho d\Omega = 0$$

After incorporating the boundary conditions, we obtain the weak form of the Poisson's equation.

$$\int_{\Gamma_1} \omega \varepsilon E_0 d\Gamma - \int_{\Omega_1} \varepsilon \vec{\nabla} \omega \cdot \vec{\nabla} \varphi d\Omega + \int_{\Omega_1} \omega \rho d\Omega = 0$$

We discretize the equation using the Galerkin method.

$$\omega = N_i, \quad i = 1, 2, ..., n$$

$$\varphi(x, y, z; t) = \sum_{j=1}^{n} N_j(x, y, z) \cdot \varphi_j(t)$$

Here, *n* represents the degrees of freedom of the finite element. Combining ε and ρ, we obtain:

$$\int_{\Omega_1} \sum_{j=1}^{n} \left( \vec{\nabla} N_i \cdot \vec{\nabla} N_j \varphi_j \right) d\Omega = \int_{\Omega_1} N_i \frac{\rho}{\varepsilon} d\Omega + \int_{\Gamma_1} N_i E_0 d\Gamma$$

The matrix representation of the discretized Poisson's equation is as follows:

$$\mathbf{M} \cdot \mathbf{\Phi} = \mathbf{f}$$

Here,

$$M_{ij} = \int_{\Omega_1} \vec{\nabla} N_i \cdot \vec{\nabla} N_j d\Omega$$

$$f_i = \int_{\Omega_1} N_i \frac{\rho}{\varepsilon} d\Omega + \int_{\Gamma_1} N_i E_0 d\Gamma$$

**Appendix G. Discretization of heat balance equations in two-temperature model**

We employ the same strategy as the Poisson equation to discretize the two-temperature heat conduction equations (see Eqs. (23) and (24)) by multiplying it with weight functions within the domain $\Omega$.

$$\int_{\Omega} \omega_e \vec{\nabla} \cdot \left( \kappa_e \vec{\nabla} T_e \right) d\Omega + \int_{\Omega} \left( P_J - G_{ep}(T_e - T_p) - C_e \frac{\partial T_e}{\partial t} \right) d\Omega = 0$$

$$\int_{\Omega} \omega_p \vec{\nabla} \cdot \left( \kappa_p \vec{\nabla} T_p \right) d\Omega + \int_{\Omega} \left( G_{ep}(T_e - T_p) - C_p \frac{\partial T_p}{\partial t} \right) d\Omega = 0$$

According to the properties of differentiation and the divergence theorem:

$$\oint_{\Gamma} \omega_e \kappa_e \vec{\nabla} T_e \cdot \vec{n} d\Gamma - \int_{\Omega} \vec{\nabla} \omega_e \cdot \kappa_e \vec{\nabla} T_e d\Omega + \int_{\Omega} \omega_e \left( P_J - G_{ep}(T_e - T_p) - C_e \frac{\partial T_e}{\partial t} \right) d\Omega = 0$$

$$\oint_{\Gamma} \omega_p \kappa_p \vec{\nabla} T_p \cdot \vec{n} d\Gamma - \int_{\Omega} \vec{\nabla} \omega_p \cdot \kappa_p \vec{\nabla} T_p d\Omega + \int_{\Omega} \omega_p \left( G_{ep}(T_e - T_p) - C_p \frac{\partial T_p}{\partial t} \right) d\Omega = 0$$

By applying the boundary conditions, the weak form is obtained:

$$\int_{\Omega_2} \omega_e C_e \frac{\partial T_e}{\partial t} d\Omega + \int_{\Omega_2} \kappa_e \vec{\nabla} \omega_e \cdot \vec{\nabla} T_e d\Omega = \int_{\Omega_2} \omega_e \left( P_J - G_{ep}(T_e - T_p) \right) d\Omega + \int_{\Gamma_3} \omega_e P_N d\Gamma$$

$$\int_{\Omega_2} \omega_p C_p \frac{\partial T_p}{\partial t} d\Omega + \int_{\Omega_2} \kappa_p \vec{\nabla} \omega_p \cdot \vec{\nabla} T_p d\Omega = \int_{\Omega_2} \omega_p G_{ep}(T_e - T_p) d\Omega$$

By equating the weight functions with the shape functions and expanding the temperature as a linear combination of them allows us to separate the temporal and spatial components of the temperature:

$$\int_{\Omega_2} C_e N_i \sum_{j=1}^{n} \left( N_j \frac{\partial T_{ej}}{\partial t} \right) d\Omega + \int_{\Omega_2} \kappa_e \sum_{j=1}^{n} \left( \vec{\nabla} N_i \cdot \vec{\nabla} N_j T_{ej} \right) d\Omega = \int_{\Omega_2} N_i \left( P_J - G_{ep}(T_e - T_p) \right) d\Omega + \int_{\Gamma_3} N_i P_N d\Gamma$$

$$\int_{\Omega_2} C_p N_i \sum_{j=1}^{n} \left( N_j \frac{\partial T_{pj}}{\partial t} \right) d\Omega + \int_{\Omega_2} \kappa_p \sum_{j=1}^{n} \left( \vec{\nabla} N_i \cdot \vec{\nabla} N_j T_{pj} \right) d\Omega = \int_{\Omega_2} N_i G_{ep}(T_e - T_p) d\Omega$$

In matrix form, we can express them as:

$$\mathbf{C}_e \cdot \frac{\partial \mathbf{T}_e}{\partial t} + \mathbf{K}_e \cdot \mathbf{T}_e = \mathbf{f}_e$$

$$\mathbf{C}_p \cdot \frac{\partial \mathbf{T}_p}{\partial t} + \mathbf{K}_p \cdot \mathbf{T}_p = \mathbf{f}_p$$

where

$$C_{eij} = \int_{\Omega_2} C_e N_i N_j d\Omega$$

$$C_{pij} = \int_{\Omega_2} C_p N_i N_j d\Omega$$

$$K_{eij} = \int_{\Omega_2} \kappa_e \vec{\nabla} N_i \cdot \vec{\nabla} N_j d\Omega$$

$$K_{pij} = \int_{\Omega_2} \kappa_p \vec{\nabla} N_i \cdot \vec{\nabla} N_j d\Omega$$

$$f_{ei} = \int_{\Omega_2} N_i \left( P_J - G_{ep} \left( T_e - T_p \right) \right) d\Omega + \int_{\Gamma_3} N_i P_N d\Gamma$$

$$f_{pi} = \int_{\Omega_2} N_i G_{ep} \left( T_e - T_p \right) d\Omega$$